\documentclass[prd,aps,showpacs,epsf,floats,10pt]{revtex4}%
\usepackage{amssymb}
\usepackage{amsfonts}
\usepackage{amsmath}
\usepackage{graphicx}%
\providecommand{\U}[1]{\protect\rule{.1in}{.1in}}
\begin{document}
\title{\textbf{Complexity and efficiency of minimum entropy production probability
paths from quantum dynamical evolutions}}
\author{\textbf{Carlo Cafaro}$^{1}$, \textbf{Shannon Ray}$^{2}$, and \textbf{Paul
M.\ Alsing}$^{2}$}
\affiliation{$^{1}$SUNY Polytechnic Institute, Albany, New York 12203, USA}
\affiliation{$^{2}$Air Force Research Laboratory, Information Directorate, Rome, New York
13441, USA}

\begin{abstract}
We present an information geometric characterization of quantum driving
schemes specified by $\mathrm{su}\left(  2\text{; }%
\mathbb{C}
\right)  $ time-dependent Hamiltonians in terms of both complexity and
efficiency concepts. Specifically, starting from pure output quantum states
describing the evolution of a spin-$1/2$ particle in an external
time-dependent magnetic field, we consider the probability paths emerging from
the parametrized squared probability amplitudes of quantum origin. The
information manifold of such paths is equipped with a Riemannian metrization
specified by the Fisher information evaluated along the parametrized squared
probability amplitudes. By employing a minimum action principle, the optimum
path connecting initial and final states on the manifold in finite-time is the
geodesic path between the two states. In particular, the total entropy
production that occurs during the transfer is minimized along these optimum
paths. For each optimum path that emerges from the given quantum driving
scheme, we evaluate the so-called information geometric complexity (IGC) and
our newly proposed \ measure of entropic efficiency constructed in terms of
the constant entropy production rates that specify the entropy minimizing
paths being compared. From our analytical estimates of complexity and
efficiency, we provide a relative ranking among the driving schemes being
investigated. Moreover, we determine that the efficiency and the temporal rate
of change of the IGC are monotonic decreasing and increasing functions,
respectively, of the constant entropic speed along these optimum paths. Then,
after discussing the connection between thermodynamic length and IGC in the
physical scenarios being analyzed, we briefly examine the link between IGC and
entropy production rate. Finally, we conclude by commenting on the fact that
an higher entropic speed in quantum transfer processes seems to necessarily go
along with a lower entropic efficiency together with a higher information
geometric complexity.

\end{abstract}

\maketitle

\section{Introduction}

The goodness of an algorithm can be assessed by a variety of criteria
\cite{kronsjo87}. In general, to quantify the performance of algorithms in
both classical and quantum settings, one considers the asymptotic scaling of a
complexity measure such as runtime or space usage with problem size
\cite{montanaro16}. Runtime is measured by the number of elementary operations
employed by the algorithm. In particular, being in the framework of quantum
computing \cite{nielsen}, runtime can be specified in terms of the number of
quantum gates applied to qubits in a quantum circuit model \cite{cafaro11}. In
general, when ranking the performances of various algorithms that solve the
same task, one usually considers the asymptotic behavior in the problem size
of the time or space complexity of the algorithm. The choice of focusing on
the asymptotic behavior is dictated by the fact that for small input sizes,
almost any algorithm can be sufficiently efficient. Addressing questions
concerning the computational performance of algorithms can be rather tricky.
For instance, specifying how long it takes to the algorithm to produce the
desired output or how much memory it needs to generate it can depend on a
number of factors, including the speed of the computer, the programming
language, the efficiency of the implementation, and the value of the input.
What does \textquotedblleft efficient\textquotedblright\ mean, exactly? Do
efficient algorithms exhibit a lower degree of complexity? Which type of
complexity are we referring to? A partial list of complexity measures we may
be making reference to includes conceptual complexity, computational
complexity, space complexity, and time complexity. It is possible to propose
efficiency measures that capture different aspects of the algorithm. For
instance, Traub's efficiency index $\eta_{\mathrm{Traub}}\overset{\text{def}%
}{=}p/\epsilon$ is an asymptotic estimate of the efficiency of an iterative
method for computing a simple real root of an $n$th degree polynomial. This
index depends on the order $p$ of convergence to the solution and on the
complexity parameter $\epsilon$ denoting the number of function evaluations
per iteration. Clearly, one may think of proposing alternative efficiency
measures, including one that takes into account the number of logical
operations performed during the algorithm as proposed by Kung and Traub in
Ref. \cite{kung73}. For an overview of different efficiency measures for
distinct iterative methods, we refer to Ref. \cite{kronsjo87}.

In recent years, numerous investigations have been carried out with the goal
of providing physical insights from Riemannian geometric characterizations
\cite{amari,ruppeiner95,ruppeiner96} of thermodynamical concepts such as
entropy production and efficiency \cite{hasegawa21,miller20,saito20,ito20}. In
Ref. \cite{hasegawa21}, using the notions of thermodynamic length,
thermodynamic divergence, and entropy production rate, the authors obtained
geometric lower bounds on the entropy production in reversible quantum
Markovian systems specified by master equations. In Ref. \cite{miller20},
making extensive use of thermodynamic geometry \cite{ruppeiner95}, the authors
presented a general technique for optimizing the thermodynamic efficiency in
microscopic quantum heat engines working close to equilibrium. In Ref.
\cite{saito20}, employing solely thermodynamic geometry arguments, the authors
found a universal trade-off between efficiency and power for microscopic
quantum heat engines driven by arbitrary periodic temperature changes. In Ref.
\cite{ito20}, relying heavily on information geometric techniques
\cite{amari}, the authors proposed an information geometric interpretation of
the entropy production for a total system and the partial entropy productions
for subsystems. Furthermore, spin models were used in \cite{ito20} to explain
in an analytical fashion these physical findings of information geometric origin.

In this paper, building upon our previous results reported in
\cite{cafaro18a,cafaro20,gassner21} and inspired by the findings uncovered in
\cite{hasegawa21,miller20,saito20,ito20}, we provide a quantitative link
between the concepts of \emph{information geometric complexity} and
\emph{entropic efficiency }by studying the entropic dynamics on information
manifolds emerging from exactly solvable time-dependent two-level quantum
systems that mimic quantum search Hamiltonians. Our motivation for considering
this type of work can be explained by pointing out a number of previous
results our proposed analysis relies on. First, there is our previous
investigation carried out in Ref. \cite{alsing19} concerning the physical
connection between quantum search Hamiltonians and exactly solvable
time-dependent two-level quantum systems \cite{messina14,grimaudo18}. Second,
there are our previous attempts in trying to provide an information geometric
perspective on the characterization of trade-offs between speed and
thermodynamic efficiency in quantum search algorithms
\cite{cafaro18a,cafaro20,gassner21}. Unfortunately, despite the agreement on
the importance that quantum algorithms should be fast and thermodynamically
efficient \cite{castelvecchi17}, there does not exist, to the best of our
knowledge, any unifying theoretical description on this matter. Our work here
aims at being a nontrivial step forward in this direction.

We provide an information geometric analysis of quantum driving schemes
characterized by $\mathrm{su}\left(  2\text{; }%
\mathbb{C}
\right)  $ time-dependent Hamiltonians by means of both complexity and
efficiency concepts. From the knowledge of the pure output quantum states
specifying the evolution of a spin-$1/2$ particle in an external magnetic
field, we construct the probability paths emerging from the parametrized
squared\textbf{ }probability amplitudes. The Fisher information evaluated
along the parametrized squared probability amplitudes provides a a Riemannian
metrization for such information manifolds. Imposing a minimum action
principle, it happens that the optimum path connecting initial and final
states on the manifold in finite-time is the geodesic path between the two
states. In particular, the total entropy production that occurs during the
quantum transfer is minimized along these optimum paths. For each optimum path
that arises from the given quantum driving Hamiltonian, we compute the
so-called information geometric complexity (IGC) and our newly proposed
\ measure of entropic efficiency. The latter quantity is expressed in terms of
the constant entropy production rates that characterize the entropy minimizing
paths being examined. From our calculations of complexity and efficiency, we
give a relative ranking among the driving schemes being compared. Moreover, we
show that the efficiency and the temporal rate of change of the IGC are
monotonic decreasing and increasing functions, respectively, of the constant
entropic speed along these optimum paths. Then, after elaborating on the
connection between thermodynamic length and IGC, we briefly discuss the
relation between IGC and entropy production rate. Finally, we conclude by
providing some remarks on the fact that an higher entropic speed in quantum
transfer processes appears to necessarily lead to a lower entropic efficiency
together with a higher IGC.

The layout of the remainder of this paper is as follows. In Section II, we
present the IGC concept. In Section III, after introducing the concepts of
thermodynamic length and thermodynamic divergence, we propose our measure of
entropic efficiency. In Section IV, we describe the quantum driving schemes
being studied and explain how to generate probability paths from the output
quantum pure state emerging from the quantum mechanical evolution. Then,
having identified the elapsed time as the key statistical parameter
\cite{braunstein96}, we apply our proposed information geometric theoretical
construct to four distinct quantum mechanical driving scenarios in Section V.
Our final remarks appear in Section VI. Finally, technical details are located
in Appendices A, B, C, and D.

\section{Information Geometric Complexity}

In this section, we introduce the concepts of information geometric entropy
(IGE) and IGC.

The IGE is a measure of complexity that was originally introduced in Ref.
\cite{cafaro07} in the context of the Information Geometric Approach to Chaos
(IGAC) theoretical setting developed in Ref. \cite{cafarothesis}. For brevity
and readability of the paper, we do not mention any superfluous detail on the
IGAC. However, for the interested reader we suggest considering the concise
discussion on the IGAC in Ref. \cite{ali18}. In what follows, we present the
concept of IGE.

Suppose that the points $\left\{  p\left(  x;\theta\right)  \right\}  $ of an
$n$-dimensional curved statistical manifold $\mathcal{M}_{s}$ are parametrized
in terms of $n$ real valued variables $\left(  \theta^{1}\text{,..., }%
\theta^{n}\right)  $, where%
\begin{equation}
\mathcal{M}_{s}\overset{\text{def}}{=}\left\{  p\left(  x;\theta\right)
:\theta=\left(  \theta^{1}\text{,..., }\theta^{n}\right)  \in\mathcal{D}%
_{\boldsymbol{\theta}}^{\mathrm{tot}}\right\}  \text{.}%
\end{equation}
The microvariables $x$ belong to the microspace $\mathcal{X}$ while the
macrovariables $\theta$ are elements of the parameter space $\mathcal{D}%
_{\boldsymbol{\theta}}^{\mathrm{tot}}$ defined as,
\begin{equation}
\mathcal{D}_{\boldsymbol{\theta}}^{\mathrm{tot}}\overset{\text{def}}{=}\left(
\mathcal{I}_{\theta^{1}}\otimes\mathcal{I}_{\theta^{2}}\text{...}%
\otimes\mathcal{I}_{\theta^{n}}\right)  \subseteq\mathbb{R}^{n}\text{.}
\label{dtot}%
\end{equation}
The quantity $\mathcal{I}_{\theta^{j}}$ in $\mathcal{D}_{\boldsymbol{\theta}%
}^{\mathrm{tot}}$ is a subset of $\mathbb{R}^{n}$ and specifies the range of
allowable values for the statistical macrovariables $\theta^{j}$. The IGE is
proposed as a measure of temporal complexity of geodesic paths within the
IGAC. The IGE\ is defined as,
\begin{equation}
\mathcal{S}_{\mathcal{M}_{s}}\left(  \tau\right)  \overset{\text{def}}{=}%
\log\widetilde{vol}\left[  \mathcal{D}_{\boldsymbol{\theta}}\left(
\tau\right)  \right]  \text{,} \label{IGE}%
\end{equation}
where the average dynamical statistical volume\textbf{\ }$\widetilde
{vol}\left[  \mathcal{D}_{\boldsymbol{\theta}}\left(  \tau\right)  \right]  $
is,
\begin{equation}
\widetilde{vol}\left[  \mathcal{D}_{\boldsymbol{\theta}}\left(  \tau\right)
\right]  \overset{\text{def}}{=}\frac{1}{\tau}\int_{0}^{\tau}vol\left[
\mathcal{D}_{\boldsymbol{\theta}}\left(  \tau^{\prime}\right)  \right]
d\tau^{\prime}\text{.} \label{rhs}%
\end{equation}
We emphasize that $\mathcal{D}_{\boldsymbol{\theta}}\left(  \tau^{\prime
}\right)  $ in Eq. (\ref{rhs}) is an $n$-dimensional subspace of
$\mathcal{D}_{\boldsymbol{\theta}}^{\mathrm{tot}}\subseteq\mathbb{R}^{n}$
whose elements $\left\{  \theta\right\}  $ with $\theta=\left(  \theta
^{1}\text{,..., }\theta^{n}\right)  $ are such that $\theta^{j}\left(
\tau_{0}\right)  \leq\theta^{j}\leq\theta^{j}\left(  \tau_{0}+\tau^{\prime
}\right)  $ with $\tau_{0}$ being the initial value assumed by the affine
parameter that specifies the geodesic paths as will be explained in more
detail shortly. Observe that the operation of temporal average is denoted with
the tilde symbol in Eq. (\ref{rhs}). For clarity, we underline the fact
that\textbf{ }$\widetilde{vol}\left[  \mathcal{D}_{\boldsymbol{\theta}}\left(
\tau\right)  \right]  $\textbf{ }in Eq. (\ref{rhs}) is defined in terms of two
sequential integration procedures. A first integration occurs on on the
explored parameter space and yields\textbf{ }$vol\left[  \mathcal{D}%
_{\boldsymbol{\theta}}\left(  \tau^{\prime}\right)  \right]  $. Then, as
second integration specifying a temporal averaging procedure is performed over
the duration of the process and leads ultimately\textbf{ }to\textbf{
}$\widetilde{vol}\left[  \mathcal{D}_{\boldsymbol{\theta}}\left(  \tau\right)
\right]  $. Moreover, the volume\textbf{\ }$vol\left[  \mathcal{D}%
_{\boldsymbol{\theta}}\left(  \tau^{\prime}\right)  \right]  $\textbf{\ }in
the RHS of Eq. (\ref{rhs}) specifies the volume of an extended region on the
manifold $\mathcal{M}_{s}$. It is defined as,
\begin{equation}
vol\left[  \mathcal{D}_{\boldsymbol{\theta}}\left(  \tau^{\prime}\right)
\right]  \overset{\text{def}}{=}\int_{\mathcal{D}_{\boldsymbol{\theta}}\left(
\tau^{\prime}\right)  }\rho\left(  \theta^{1}\text{,..., }\theta^{n}\right)
d^{n}\theta\text{.} \label{v}%
\end{equation}
The quantity $\rho\left(  \theta^{1}\text{,..., }\theta^{n}\right)
\overset{\text{def}}{=}\sqrt{g\left(  \theta\right)  }$ is the so-called
Fisher density and is equal to the square root of the determinant $g\left(
\theta\right)  $ of the Fisher-Rao information metric tensor $g_{ij}\left(
\theta\right)  $, $g\left(  \theta\right)  \overset{\text{def}}{=}\det\left[
g_{ij}\left(  \theta\right)  \right]  $. The quantity $g_{ij}\left(
\theta\right)  $ is given by
\begin{equation}
g_{ij}\left(  \theta\right)  \overset{\text{def}}{=}\int p\left(
x|\theta\right)  \partial_{i}\log p\left(  x|\theta\right)  \partial_{j}\log
p\left(  x|\theta\right)  dx\text{,} \label{FRmetric}%
\end{equation}
with $\partial_{i}\overset{\text{def}}{=}\partial/\partial\theta^{i}$. The
expression of $vol\left[  \mathcal{D}_{\boldsymbol{\theta}}\left(
\tau^{\prime}\right)  \right]  $ in Eq. (\ref{v}) becomes more transparent for
manifolds with information metric tensor whose determinant can be factorized
as
\begin{equation}
g\left(  \theta\right)  =g\left(  \theta^{1}\text{,..., }\theta^{n}\right)
={\prod\limits_{j=1}^{n}}g_{j}\left(  \theta^{j}\right)  \text{.}%
\end{equation}
In this case, the IGE in Eq. (\ref{IGE}) can be recast as
\begin{equation}
\mathcal{S}_{\mathcal{M}_{s}}\left(  \tau\right)  =\log\left\{  \frac{1}{\tau
}\int_{0}^{\tau}\left[  {\prod\limits_{j=1}^{n}}\left(  \int_{\tau_{0}}%
^{\tau_{0}+\tau^{\prime}}\sqrt{g_{j}\left[  \theta^{j}\left(  \xi\right)
\right]  }\frac{d\theta^{j}}{d\xi}d\xi\right)  \right]  d\tau^{\prime
}\right\}  \text{.} \label{IGEmod}%
\end{equation}
We emphasize that for correlated microvariables $\left\{  x\right\}  $,
$g\left(  \theta\right)  $ is not factorizable and the general definition of
the IGE must be employed. For a discussion on the effects of microscopic
correlations on the IGE of Gaussian statistical models, we refer to Ref.
\cite{ali10}\textbf{.} Within the IGAC, the leading asymptotic behavior of
$\mathcal{S}_{\mathcal{M}_{s}}\left(  \tau\right)  $ in Eq. (\ref{IGEmod}) is
used to characterize the complexity of the statistical models being
investigated. For this purpose, we take into consideration the leading
asymptotic term in the IGE expression,
\begin{equation}
\mathcal{S}_{\mathcal{M}_{s}}^{\text{\textrm{asymptotic}}}\left(  \tau\right)
\sim\lim_{\tau\rightarrow\infty}\left[  \mathcal{S}_{\mathcal{M}_{s}}\left(
\tau\right)  \right]  \text{.} \label{LONG}%
\end{equation}
We point out that $\mathcal{D}_{\boldsymbol{\theta}}\left(  \tau^{\prime
}\right)  $ specifies the integration space that appears in the definition
of\textbf{ }$vol\left[  \mathcal{D}_{\boldsymbol{\theta}}\left(  \tau^{\prime
}\right)  \right]  $ in Eq. (\ref{v}). It is given\textbf{ }by,%
\begin{equation}
\mathcal{D}_{\boldsymbol{\theta}}\left(  \tau^{\prime}\right)  \overset
{\text{def}}{=}\left\{  \theta:\theta^{j}\left(  \tau_{0}\right)  \leq
\theta^{j}\leq\theta^{j}\left(  \tau_{0}+\tau^{\prime}\right)  \right\}
\text{,}%
\end{equation}
where $\theta^{j}=\theta^{j}\left(  \xi\right)  $ with $\tau_{0}\leq\xi
\leq\tau_{0}+\tau^{\prime}$ and $\tau_{0}$ denoting the initial value of the
affine parameter $\xi$ such that,
\begin{equation}
\frac{d^{2}\theta^{j}\left(  \xi\right)  }{d\xi^{2}}+\Gamma_{ik}^{j}%
\frac{d\theta^{i}}{d\xi}\frac{d\theta^{k}}{d\xi}=0\text{.} \label{ge}%
\end{equation}
The quantities $\Gamma_{ik}^{j}$ in\ Eq. (\ref{ge}) are the Christoffel
connection coefficients,%
\begin{equation}
\Gamma_{ik}^{j}\overset{\text{def}}{=}\frac{1}{2}g^{jl}\left(  \partial
_{i}g_{jk}+\partial_{k}g_{il}-\partial_{l}g_{ik}\right)  \text{.}%
\end{equation}
The integration domain $\mathcal{D}_{\boldsymbol{\theta}}\left(  \tau^{\prime
}\right)  $ is an $n$-dimensional subspace of $\mathcal{D}_{\boldsymbol{\theta
}}^{\mathrm{tot}}$ whose elements are $n$-dimensional macrovariables $\left\{
\theta\right\}  $ with components $\theta^{j}$ bounded by given limits of
integration $\theta^{j}\left(  \tau_{0}\right)  $ and $\theta^{j}\left(
\tau_{0}+\tau^{\prime}\right)  $. The integration of the $n$-coupled nonlinear
second order ODEs in Eq. (\ref{ge}) determines the temporal functional form of
such limits. Having defined the IGE, we call the information geometric
complexity (IGC) the quantity $\mathcal{C}_{\mathcal{M}_{s}}\left(
\tau\right)  $ defined as%
\begin{equation}
\mathcal{C}_{\mathcal{M}_{s}}\left(  \tau\right)  \overset{\text{def}}%
{=}\widetilde{vol}\left[  \mathcal{D}_{\boldsymbol{\theta}}\left(
\tau\right)  \right]  =e^{\mathcal{S}_{\mathcal{M}_{s}}\left(  \tau\right)
}\text{.} \label{IGC}%
\end{equation}
In particular, we shall focus on the asymptotic temporal behavior of the
complexity as described by $\mathcal{C}_{\mathcal{M}_{s}}%
^{\text{\textrm{asymptotic}}}\left(  \tau\right)  \overset{\tau\rightarrow
\infty}{\sim}e^{\mathcal{S}_{\mathcal{M}_{s}}\left(  \tau\right)  }$. \ 

To interpret $\mathcal{C}_{\mathcal{M}_{s}}\left(  \tau\right)  $, we simply
give an interpretation of $\mathcal{S}_{\mathcal{M}_{s}}\left(  \tau\right)
$. This latter quantity is defined in Eq. (\ref{IGE}) as an affine temporal
average of the\textbf{\ }$n$\textbf{-}fold integral of the Fisher density over
geodesic paths viewed as maximum probability trajectories and serves as a
measure of the number of the accessible macrostates in the statistical
configuration manifold\textbf{. }More specifically, the IGE at a specific
instant is defined as the logarithm of the volume of the effective parameter
space explored by the system at that very instant. We introduce the temporal
averaging procedure in Eq. (\ref{rhs}) to average out the possibly very
complex fine details of the entropic dynamical description of the system on
the underlying curved statistical manifold. Furthermore, we consider the
long-time limit in\ Eq. (\ref{LONG}) to characterize in a proper fashion the
chosen dynamical indicators of chaoticity by removing the transient effects
which enter the computation of the expected value of the volume of the
effective parameter space. Therefore,\emph{\ }the IGE is constructed to
provide an asymptotic coarse-grained inferential description of the complex
dynamics of a system in the presence of incomplete information. For further
details on the IGE and IGC, we refer to Refs.
\cite{cafaro18,cafaro17,cafaro10}.

In this paper, we focus on quantifying the IGC of parametrized probability
paths $\left\{  p_{x}\left(  \theta\right)  \right\}  $ with $p_{x}\left(
\theta\right)  \overset{\text{def}}{=}p\left(  x|\theta\right)  $ constructed
from the time-dependent transition probabilities between orthogonal initial
and final quantum states $\left\{  \left\vert w\right\rangle \text{,
}\left\vert w_{\perp}\right\rangle \right\}  $ emerging from selected quantum
mechanical evolutions of two-level quantum systems (see Section IV). The
single parameter $\theta$ used in the parametrization can be regarded as the
statistical version of the elapsed time $t$. In particular, $\theta$ is
assumed to be an experimental parameter that can be characterized by measuring
a suitable time-dependent observable quantity such as the transverse magnetic
field intensity\textbf{ }$B_{\perp}\left(  t\right)  $\textbf{.}

Having introduced the IGC concept in Eq. (\ref{IGC}), we propose our measure
of entropic efficiency in the next section.

\section{Efficiency}

In this section, after recalling the notions of thermodynamic length and
thermodynamic divergence, we propose our measure of entropic efficiency.

\subsection{Thermodynamic length and divergence}

Thermodynamic systems can be specified by Riemannian manifolds equipped with a
thermodynamic metric tensor that is identical to the Fisher information metric
\cite{crooks07}, once the theory of fluctuations is included into the axioms
of equilibrium thermodynamics \cite{ruppeiner79}. Then, this Riemannian
structure allows one to define the notion of length for fluctuations about
equilibrium states as well as for thermodynamic processes proceeding via
equilibrium states. Originally, Weinhold presented a Riemannian metric in the
space of thermodynamic equilibrium states employing the second derivatives of
the internal energy with respect to extensive variables in Ref.
\cite{weinhold75}. Subsequently, Ruppeiner proposed a Riemannian geometric
model of thermodynamics with a Riemann structure defined by means of a metric
tensor specified by second derivatives of the entropy as a function of
extensive variables (such as volume and mole number, for instance) in Ref.
\cite{ruppeiner79}. Salamon and Berry introduced the notion of thermodynamic
length by employing the energy version of the thermodynamic metric tensor
$g_{\alpha\beta}\left(  \theta\right)  $ in Ref. \cite{salamon83},%
\begin{equation}
\mathcal{L}\left(  \bar{\tau}/\tau_{\ast}\right)  \overset{\text{def}}{=}%
\int_{0}^{\bar{\tau}/\tau_{\ast}}\left(  \frac{d\theta^{\alpha}}%
{dt_{\text{\textrm{th}}}}g_{\alpha\beta}\left(  \theta\right)  \frac
{d\theta^{\beta}}{dt_{\text{\textrm{th}}}}\right)  ^{1/2}dt_{\text{\textrm{th}%
}}\text{,} \label{length1}%
\end{equation}
with $t_{\text{\textrm{th}}}$ denoting the dimensionless thermodynamic time
$t_{\text{\textrm{th}}}$ with $0\leq t_{\text{\textrm{th}}}\leq\bar{\tau}%
/\tau_{\ast}$. Furthermore, $\bar{\tau}$ and $\tau_{\ast}$ are the duration
time and the mean internal relaxation time of the physical process under
consideration, respectively. For clarity, we point out that the mean
relaxation time $\tau_{\ast}$ is an indicator of how fast the physical system
reaches an equilibrium configuration with an environment with which it is
brought into contact. In particular, to a smaller value of $\tau_{\ast}$ there
corresponds a faster equilibration of the system-environment system. For a
detailed discussion on the concepts of instantaneous and mean relaxation times
in molecular physics, we refer to Ref. \cite{feldmann85}. Upon identifying the
affine parameter $\xi$ with the the dimensionless thermodynamic time
$t_{\text{\textrm{th}}}$ and the duration of the process $\tau$ with
$\bar{\tau}/\tau_{\ast}$,\textbf{\ }the thermodynamic length in Eq.
(\ref{length1}) of a path $\gamma_{\theta}$\textbf{ }with the
parameter\textbf{ }$\theta$ parametrized by an affine parameter $\xi$ with
$0\leq\xi\leq\tau$ in the space of thermal states becomes%
\begin{equation}
\mathcal{L}\left(  \tau\right)  =\int_{0}^{\tau}\left(  \frac{d\theta^{\alpha
}}{d\xi}g_{\alpha\beta}\left(  \theta\right)  \frac{d\theta^{\beta}}{d\xi
}\right)  ^{1/2}d\xi\text{.} \label{length2}%
\end{equation}
The quantity $\mathcal{L}\left(  \tau\right)  $ in\ Eq. (\ref{length2}) is
measured by the number of natural fluctuations along the path $\gamma_{\theta
}$. The larger the fluctuations, the closer the points are together. Indeed,
in analogy to Wootters' statistical distance between probability distributions
\cite{wootters81}, the thermodynamic length can be interpreted as a measure of
the maximal number of statistically distinguishable thermodynamic states along
the path $\gamma_{\theta}$ \cite{diosi84}. As a matter of fact, following
Wootters, we can interpret the points $\theta$ and $\theta+d\theta$ long the
path $\gamma_{\theta}$ as statistically distinguishable if $d\theta$ is at
least equal to the standard fluctuation of $\theta$. In terms of the distance
$ds^{2}=g_{\alpha\beta}\left(  \theta\right)  d\theta^{\alpha}d\theta^{\beta}%
$, this is equivalent to $ds^{2}\geq1$. Clearly, $\mathcal{L}\left(
\tau\right)  $ in\ Eq. (\ref{length2}) has dimensions of $(\mathrm{energy}%
)^{1/2}$ if one uses the energy version of the thermodynamic metric tensor.
If, instead, one uses the entropy version of the thermodynamic metric tensor,
$\mathcal{L}\left(  \tau\right)  $ has dimensions of $(\mathrm{entropy}%
)^{1/2}$. To better understand the physical interpretation of the
thermodynamic length, it is helpful to introduce the so-called thermodynamic
divergence $\mathcal{I}\left(  \tau\right)  $ of a path $\gamma_{\theta}$ with
the variable\textbf{ }$\theta$\textbf{ }expressed in terms of an affine
parameter $\xi$ with $0\leq\xi\leq\tau$ as in Eq. (\ref{length2}),%
\begin{equation}
\mathcal{I}\left(  \tau\right)  \overset{\text{def}}{=}\int_{0}^{\tau}%
\frac{d\theta^{\alpha}}{d\xi}g_{\alpha\beta}\left(  \theta\right)
\frac{d\theta^{\beta}}{d\xi}d\xi\text{.} \label{divergence}%
\end{equation}
The quantity $\mathcal{I}\left(  \tau\right)  $ in Eq. (\ref{divergence}) is a
measure of the losses (or, dissipation) in the process quantified by the total
entropy produced (or, dissipated availability \cite{salamon83}) along the path
$\gamma_{\theta}$. Applying the Cauchy-Schwarz inequality with integrals of
functions,%
\begin{equation}
\left[  \int_{0}^{\tau}f_{1}^{2}\left(  \xi\right)  d\xi\right]  \left[
\int_{0}^{\tau}f_{2}^{2}\left(  \xi\right)  d\xi\right]  \geq\left[  \int
_{0}^{\tau}f_{1}\left(  \xi\right)  f_{2}\left(  \xi\right)  d\xi\right]
^{2}\text{,}%
\end{equation}
and using Eqs. (\ref{length2}) and (\ref{divergence}), it happens that
$\mathcal{I}\geq\tau^{-1}\mathcal{L}^{2}$ with $\tau\overset{\text{def}}%
{=}\bar{\tau}/\tau_{\ast}$ once we identify $f_{1}\left(  \xi\right)  $ and
$f_{2}\left(  \xi\right)  $ with $ds/d\xi$ and\textbf{ }$1$\textbf{,
}respectively\textbf{.} Therefore, the square of the thermodynamic length of
the path $\gamma_{\theta}$ multiplied by the ratio of the internal relaxation
time of the system to the duration of the process furnishes a lower bound to
the dissipation in the process. This bound is more realistic than the (ideal)
reversible bound which would be equal to zero. The equality $\mathcal{I}%
=\mathcal{I}_{\min}\overset{\text{def}}{=}\tau^{-1}\mathcal{L}^{2}$ is
obtained when the thermodynamic speed is constant along the path
$\gamma_{\theta}$. Therefore, the process exhibits minimum losses when it
produces minimum entropy. This happens when it proceeds at constant speed,
with the entropy production rate being equal to the squared thermodynamic
speed itself.

Let $n_{\mathcal{M}_{s}}$ be the dimensionality of the parameter space with
$\theta\left(  \xi\right)  \overset{\text{def}}{=}\left\{  \theta^{\alpha
}\left(  \xi\right)  \right\}  _{1\leq\alpha\leq n_{\mathcal{M}_{s}}}$ and
$0\leq\xi\leq\tau$. Then, the optimum paths $\gamma_{\theta}$ are paths
characterized by the most favorable affine time $\xi$ parametrization yielding
the shortest thermodynamic length. More explicitly, the optimum paths satisfy
the geodesic equation that can be obtained via variational calculus by
minimizing the action functional represented by the thermodynamic length in
Eq. (\ref{length2}). One imposes that $\delta\mathcal{L}$ is equal to zero
subject to the constraint that $\delta\theta^{\alpha}=0$ at the extremum. We
point out that $\xi$ is defined up to changes of scale and origin and, thus,
is not unique. Interestingly, we emphasize that the optimum paths that
minimize $\mathcal{L}\left(  \tau\right)  $ in\ Eq. (\ref{length2}) are the
paths that minimize the divergence $\mathcal{I}\left(  \tau\right)  $ in\ Eq.
(\ref{divergence}). As a matter of fact, minimizing $\mathcal{I}\left(
\tau\right)  $ under the same working conditions used in the minimization of
$\mathcal{L}\left(  \tau\right)  $, it happens that the optimum paths
$\theta^{\alpha}\left(  \xi\right)  $ satisfy the equation%
\begin{equation}
\frac{d}{d\xi}\left[  g_{\alpha\rho}\left(  \theta\right)  \frac
{d\theta^{\alpha}}{d\xi}\right]  -\frac{1}{2}\frac{d\theta^{\alpha}}{d\xi
}\frac{\partial g_{\alpha\beta}\left(  \theta\right)  }{\partial\theta^{\rho}%
}\frac{d\theta^{\beta}}{d\xi}=0\text{.} \label{general1}%
\end{equation}
It is worth noting that Eq. (\ref{general1}) is the information geometric
analogue of Eqs. $(36)$ and $(6)$ in Refs. \cite{diosi96} and \cite{crooks17},
respectively. For an explicit verification of the interchangeability between
the geodesic equations emerging from the variations of $\delta\left(
\int\sqrt{ds^{2}}\right)  $ and $\delta\left(  \int ds^{2}\right)  $, we refer
to Appendix A. Since optimum paths are geodesic paths, the \textquotedblleft
thermodynamic\textquotedblright\ speed is constant when evaluated along these
shortest paths. Henceforth, we shall name this speed \textquotedblleft
entropic\textquotedblright\ speed $v_{\text{\textrm{E}}}$ and define it as%
\begin{equation}
v_{\text{\textrm{E}}}\overset{\text{def}}{=}\left[  \frac{d\theta^{\alpha}%
}{d\xi}g_{\alpha\beta}\left(  \theta\right)  \frac{d\theta^{\beta}}{d\xi
}\right]  ^{1/2}\text{.} \label{vthermo}%
\end{equation}
Moreover, optimum paths are also paths specified by constant entropy
production rate $r_{\text{\textrm{E}}}$ (that is, the squared invariant norm
of the speed $v_{\text{\textrm{E}}}$), with $r_{\text{\textrm{E}}}$ given by%
\begin{equation}
r_{\mathrm{E}}\overset{\text{def}}{=}\frac{d}{d\tau}\mathcal{I}(\tau)=\frac
{d}{d\tau}\left[  \int_{0}^{\tau}\frac{d\theta^{\alpha}}{d\xi}g_{\alpha\beta
}\left(  \theta\right)  \frac{d\theta^{\beta}}{d\xi}d\xi\right]  \text{,}
\label{EPR}%
\end{equation}
with the thermodynamic divergence $\mathcal{I}\left(  \tau\right)  $ defined
in Eq. (\ref{divergence}) and evaluated along the optimum paths. For clarity,
we stress that we are interested here in the global (i.e., integral) problem
of minimizing the entropy production over the complete path.\ Alternatively,
one may be interested in the local (i.e., differential) problem of minimizing
the rate of entropy dissipation at each instant of time \cite{andresen94}.
Moreover, for completeness, we point out that both minimum entropy production
and constant entropy production rate occur along geodesic paths in
thermodynamic state space for optimal (linear) processes with $g_{\alpha\beta
}=g_{\alpha\beta}\left(  \theta\right)  $. For a discussion on the non
constancy of the rate of entropy production within the framework of
non-linearized thermodynamics of irreversible processes with $g_{\alpha\beta
}=g_{\alpha\beta}\left(  \theta\text{, }\dot{\theta}\right)  $, we refer to
\cite{spirkl95}.

To better grasp the physical interpretation of $r_{\mathrm{E}}$ in Eq.
(\ref{EPR}), we note two facts. First, the thermodynamic metric tensor
$g_{\alpha\beta}\left(  \theta\right)  $ equals $\overline{\delta X}%
_{\alpha\beta}^{2}$, with%
\begin{equation}
\overline{\delta X}_{\alpha\beta}^{2}\overset{\text{def}}{=}\left\langle
\left(  X_{\alpha}-\left\langle X_{\alpha}\right\rangle \right)  \left(
X_{\beta}-\left\langle X_{\beta}\right\rangle \right)  \right\rangle \text{.}
\label{covid}%
\end{equation}
The quantity $\overline{\delta X}_{\alpha\beta}^{2}$ in Eq. (\ref{covid}) is
the covariance matrix of fluctuations around equilibrium defined in terms of
the thermodynamic variables $\left\{  X_{\alpha}\left(  x\right)  \right\}  $
that characterize the Hamiltonian of the system. The quantity $\left\{
x\right\}  $ denotes the set of relevant configuration space variables.
Second, consider the canonical Gibbs distribution function $p\left(
x|\theta\right)  \equiv p_{x}\left(  \theta\right)  $ with $p_{x}\left(
\theta\right)  $ defined as\textbf{ }%
\begin{equation}
p_{x}\left(  \theta\right)  \overset{\text{def}}{=}\frac{e^{-\theta^{\alpha
}\left(  \xi\right)  X_{\alpha}\left(  x\right)  }}{\mathcal{Z}}%
\mathbf{\ }\text{,} \label{plug}%
\end{equation}
with\textbf{ }$\mathcal{Z}$\textbf{ }being the partition function of the
system. Inserting $p_{x}\left(  \theta\right)  $ in Eq. (\ref{plug}) into the
usual definition of the Fisher-Rao information metric tensor $g_{\alpha\beta
}\left(  \theta\right)  $, it can be shown that this latter quantity equals
the thermodynamic metric tensor. In other words, $g_{\alpha\beta}\left(
\theta\right)  $ is equal to $\overline{\delta X}_{\alpha\beta}^{2}$ in Eq.
(\ref{covid}). Then, a simple calculation yields the following alternative
expression of $r_{\mathrm{E}}$ in Eq. (\ref{EPR}),%
\begin{equation}
r_{\mathrm{E}}=\frac{d\theta^{\alpha}}{d\xi}\overline{\delta X}_{\alpha\beta
}^{2}\frac{d\theta^{\beta}}{d\xi}=%
{\displaystyle\sum\limits_{x}}
p_{x}(\theta)\left(  \frac{d\log p_{x}(\theta)}{d\xi}\right)  ^{2}\text{,}
\label{er3}%
\end{equation}
Therefore, $r_{\mathrm{E}}$ in Eq. (\ref{er3}) can be also described as the
\textquotedblleft product\textquotedblright\ of the fluctuation term
$\overline{\delta X}_{\alpha\beta}^{2}$ and the square of the total rate of
change with respect to the affine parameter $\xi$ of the control parameter
$\theta^{\alpha}\left(  \xi\right)  $. Note that in heat transfer problems, be
it cooling or heating, the control parameter is given by temperature. However,
in mass transfer problems, in magnetic systems, and in elastic systems,
suitable control parameters are specified by chemical potential, magnetic
field, and stress, respectively. For a more detailed discussion on the
physical significance of the concept of entropy production rate in relation to
the thermodynamics of a system of spin-$1/2$ particles driven by an external
magnetic field, we refer to Appendix B. For the sake of forthcoming
discussions, we shall be naming lengths, divergences, and speeds as
\textquotedblleft entropic\textquotedblright\ quantities.

\subsection{Entropic efficiency}

In what follows, we would like to propose an efficiency measure $\eta
_{\mathrm{E}}$ with $0\leq$ $\eta_{\mathrm{E}}\leq1$ for the various driving
schemes in terms of the rate of entropy production $r_{\mathrm{E}}$ along the
path $\gamma_{\theta}$.

In Ref. \cite{cafaro20}, we proposed an asymmetric efficiency measure
$\eta_{\mathrm{E}}^{\left(  1\right)  }$ where the hottest path corresponded
to the least efficient driving scheme. The efficiency $\eta_{\mathrm{E}%
}^{\left(  1\right)  }$ was defined as%
\begin{equation}
\eta_{\mathrm{E}}^{\left(  1\right)  }\left(  r_{\mathrm{E}}\right)
\overset{\text{def}}{=}1-\frac{r_{\mathrm{E}}}{r_{\mathrm{E}}^{\max}}\text{,}
\label{eff1}%
\end{equation}
where $0\leq\eta_{\mathrm{E}}^{\left(  1\right)  }\left(  r_{\mathrm{E}%
}\right)  \leq1$ for any $0\leq r_{\mathrm{E}}\leq r_{\mathrm{E}}^{\max}$ with
$\eta_{\mathrm{E}}^{\left(  1\right)  }\left(  r_{\mathrm{E}}^{\max}\right)
=0$. This efficiency was partially inspired by the definition of thermal
efficiency of a heat engine \cite{beretta05} and by the notion of efficiency
of a quantum evolution in the Riemannian approach to quantum mechanics as
presented in Refs. \cite{anandan90,cafaroPRA20}.\ The thermal efficiency
$\eta_{\text{thermo}}$ of a heat engine in thermodynamics can be defined as
\begin{equation}
\eta_{\text{\textrm{thermo}}}\overset{\text{def}}{=}1-\frac{Q_{\text{out}}%
}{Q_{\text{in}}}\text{,} \label{efftermo}%
\end{equation}
with $Q_{\text{out}}$ and $Q_{\text{in}}$ being the output and input thermal
energies with $W_{\text{out}}\overset{\text{def}}{=}Q_{\text{in}%
}-Q_{\text{out}}\geq0$ denoting the actual work performed by the heat engine
\cite{beretta05}. In the Riemannian approach to quantum mechanics, instead,
the efficiency of a quantum evolution is defined as $\eta_{\text{\textrm{QM}}%
}\overset{\text{def}}{=}1-\Delta s/s$ with $0\leq\eta_{\text{QM}}\leq1$ and
$\Delta s\overset{\text{def}}{=}s-s_{0}$. The quantity $s_{0}$ represents the
dimensionless distance along the shortest geodesic path (ideal) $\gamma
_{\text{\textrm{ideal}}}$ joining the fixed initial ($\left\vert
A\right\rangle $) and final ($\left\vert B\right\rangle $) points of the
evolution that are distinct points on the complex projective Hilbert space.
The quantity $s$ instead, denotes the distance along the effective (real) path
$\gamma_{\text{\textrm{real}}}$ connecting $\left\vert A\right\rangle $ and
$\left\vert B\right\rangle $ and is measured by the Fubini-Study metric. The
quantum evolution is maximally efficient when the evolution occurs with
minimum time-energy uncertainty. This scenario is specified by $\eta
_{\text{\textrm{QM}}}=1$ and happens when $\gamma_{\text{\textrm{real}}}$ and
$s$ approach $\gamma_{\text{\textrm{ideal}}}$ and $s_{0}$, respectively.
Concerning this latter inspiration, we replaced the quantum mechanical
condition of maximum energy dispersion with the information-theoretic
requirement of minimum entropy production. Then, we found it appropriate to
propose a definition of entropic efficiency of an evolution along a path of
minimum entropic length joining the distinct initial and final points on the
information manifold as the above mentioned quantity $\eta_{\mathrm{E}%
}^{\left(  1\right)  }\left(  r_{\mathrm{E}}\right)  $. In this efficiency
definition, $r_{\mathrm{E}}^{\max}$ plays the effective role of a normalizing
factor that makes $\eta_{E}$ adimensional with $0\leq\eta_{\mathrm{E}%
}^{\left(  1\right)  }\left(  r_{\mathrm{E}}\right)  \leq1$. Then, unit
entropic efficiency can be achieved when the evolution is characterized by a
path that is maximally cooled (that is, maximally reversible). In such a case,
the total entropy production remains ideally constant during the evolution
and, as a consequence, the rate of entropy production $r_{\mathrm{E}}$
vanishes. Alternatively, one may think of proposing a different asymmetric
efficiency measure\textbf{ }$\eta_{\mathrm{E}}^{\left(  2\right)  }\left(
r_{\mathrm{E}}\right)  $ where the coolest path is the most efficient. In this
case, one can propose a measure $\eta_{\mathrm{E}}^{\left(  2\right)  }\left(
r_{\mathrm{E}}\right)  $ given by%
\begin{equation}
\eta_{\mathrm{E}}^{\left(  2\right)  }\left(  r_{\mathrm{E}}\right)
\overset{\text{def}}{=}\frac{r_{\mathrm{E}}^{\min}}{r_{\mathrm{E}}}\text{,}
\label{eff2}%
\end{equation}
where $0\leq\eta_{\mathrm{E}}^{\left(  2\right)  }\left(  r_{\mathrm{E}%
}\right)  \leq1$ for any $0\leq r_{\mathrm{E}}^{\min}\leq r_{\mathrm{E}}$ with
$\eta_{\mathrm{E}}^{\left(  2\right)  }\left(  r_{\mathrm{E}}^{\min}\right)
=1$. We point out that both measures $\eta_{\mathrm{E}}^{\left(  1\right)
}\left(  r_{\mathrm{E}}\right)  $\textbf{ }in Eq. (\ref{eff1}) and
$\eta_{\mathrm{E}}^{\left(  2\right)  }\left(  r_{\mathrm{E}}\right)  $ in Eq.
(\ref{eff2}) preserve the relative ranking of paths. In addition, they
are\textbf{ }asymmetric measures since $r_{\mathrm{E}}^{\max}$ and
$r_{\mathrm{E}}^{\min}$ play special roles in the ranking procedure. However,
in both ranking schemes, $r_{\mathrm{E}}^{\max}$ and $r_{\mathrm{E}}^{\min}$
belong to the set of entropy production rates that specify the paths being
ranked. Specifically, $r_{\mathrm{E}}^{\min}=r_{\mathrm{E}}^{(k)}$ and
$r_{\mathrm{E}}^{\max}=r_{\mathrm{E}}^{(k^{\prime})}$ belong to $\left\{
r_{\mathrm{E}}^{(i)}\right\}  _{1\leq i\leq\bar{N}}$ for some $k\neq
k^{\prime}\in\left\{  1\text{,..., }\bar{N}\right\}  $ with $\bar{N}$ denoting
the number of driving schemes being ranked. Therefore, $r_{\mathrm{E}}^{\min}$
($r_{\mathrm{E}}^{\max}$) does not represent an absolute external minimum
(maximum) to be achieved in an ideal best (worst) scenario. Moreover,
depending on the particular tuning of the parameters that specify the driving
Hamiltonian, $r_{\mathrm{E}}^{\min}$ and $r_{\mathrm{E}}^{\max}$ can change.
More explicitly, assuming the tuning of a single parameter (for instance, the
frequency of oscillation of a time-dependent external magnetic field), there
could be a range of values of this parameter for which $\left(  r_{\mathrm{E}%
}^{\min}\text{, }r_{\mathrm{E}}^{\max}\right)  =\left(  r_{\mathrm{E}}%
^{(k)}\text{, }r_{\mathrm{E}}^{(k^{\prime})}\right)  $ and a different range
for which $\left(  r_{\mathrm{E}}^{\min}\text{, }r_{\mathrm{E}}^{\max}\right)
=\left(  r_{\mathrm{E}}^{(\tilde{k})}\text{, }r_{\mathrm{E}}^{(\tilde
{k}^{\prime})}\right)  $ with $k\neq\tilde{k}$ and/or $k^{\prime}\neq\tilde
{k}^{\prime}$. Motivated by the lack of an absolute optimal driving scheme of
reference (unlike the quantum scenarios studied in Refs.
\cite{anandan90,cafaroPRA20}) and maintaining the willingness of preserving
the idea of dependence of the entropic efficiency on the rate of entropy
production (with the coolest paths being the most efficient and the hottest
paths being the least efficient), we propose in this paper a symmetric measure
of entropic efficiency given by%
\begin{equation}
\eta_{\mathrm{E}}\left(  r_{\mathrm{E}}^{\left(  l\right)  }\text{,
}r_{\mathrm{E}}^{\left(  m\right)  }\right)  \overset{\text{def}}{=}%
1-\frac{\left\vert r_{\mathrm{E}}^{\left(  l\right)  }-r_{\mathrm{E}}^{\left(
m\right)  }\right\vert }{r_{\mathrm{E}}^{\left(  l\right)  }+r_{\mathrm{E}%
}^{\left(  m\right)  }}\text{,} \label{efficiency}%
\end{equation}
with $0\leq\eta_{\mathrm{E}}\left(  r_{\mathrm{E}}^{\left(  l\right)  }\text{,
}r_{\mathrm{E}}^{\left(  m\right)  }\right)  \leq1$ by construction for any
pair of positive $r_{\mathrm{E}}^{\left(  l\right)  }$ and $r_{\mathrm{E}%
}^{\left(  m\right)  }$. Furthermore, $\eta_{\mathrm{E}}\left(  r_{\mathrm{E}%
}^{\left(  l\right)  }\text{, }r_{\mathrm{E}}^{\left(  m\right)  }\right)  $
preserves the relative ranking of paths that one obtains by means of
$\eta_{\mathrm{E}}^{\left(  1\right)  }\left(  r_{\mathrm{E}}\right)  $ and
$\eta_{\mathrm{E}}^{\left(  2\right)  }\left(  r_{\mathrm{E}}\right)  $. For
an explicit check of this conservation behavior, we refer to Appendix C.
Clearly, although preserving the relative ranking of paths, $\eta_{\mathrm{E}%
}^{\left(  1\right)  }\left(  r_{\mathrm{E}}\right)  $ and $\eta_{\mathrm{E}%
}^{\left(  2\right)  }\left(  r_{\mathrm{E}}\right)  $ assume relatively
different numerical values. For instance, while $\eta_{\mathrm{E}}^{\left(
2\right)  }\left(  r_{\mathrm{E}}\right)  \rightarrow1$ as $r_{\mathrm{E}%
}\rightarrow r_{\mathrm{E}}^{\min}$, $\eta_{\mathrm{E}}^{\left(  1\right)
}\left(  r_{\mathrm{E}}\right)  \rightarrow1$ in the extreme scenario in which
$r_{\mathrm{E}}\rightarrow0$. Moreover, while $\eta_{\mathrm{E}}^{\left(
1\right)  }\left(  r_{\mathrm{E}}\right)  \rightarrow0$ as $r_{\mathrm{E}%
}\rightarrow r_{\mathrm{E}}^{\max}$, $\eta_{\mathrm{E}}^{\left(  2\right)
}\left(  r_{\mathrm{E}}\right)  \rightarrow0$ in the extreme scenario in which
$r_{\mathrm{E}}\rightarrow\infty$. In our paper, one of the two values between
$r_{\mathrm{E}}^{\left(  l\right)  }$ and $r_{\mathrm{E}}^{\left(  m\right)
}$ (say, $r_{\mathrm{E}}^{\left(  m\right)  }$) is picked as $r_{\mathrm{E}%
}^{\min}$ for a given range of values of the Hamiltonian parameter being
tuned. Then, our proposed measure of efficiency assumes unit value when
$r_{\mathrm{E}}^{\left(  l\right)  }=r_{\mathrm{E}}^{\left(  m\right)  }$ with
$r_{\mathrm{E}}^{\left(  m\right)  }\overset{\text{def}}{=}r_{\mathrm{E}%
}^{\min}$ and tends to vanish when $r_{\mathrm{E}}^{\left(  l\right)  }\gg
r_{\mathrm{E}}^{\min}$. For a detailed physical discussion on the idea of
irreversible entropy production when analyzing the causes of inefficiency in
thermodynamic systems, we refer to Ref. \cite{tolman48}.

Having introduced the IGC in Eq. (\ref{IGC}) and the entropic efficiency in
Eq. (\ref{efficiency}), we are ready to describe the quantum driving schemes
that we study in the next section.

\section{Quantum driving schemes}

In this section, we introduce the quantum driving schemes being investigated
and mention the manner in which one can generate probability paths from the
output quantum pure state emerging from the quantum mechanical evolution.

\subsection{Probability paths from driving schemes}

Inspired by the link between analog quantum search and two-level quantum
systems \cite{alsing19,alsing19b} and following Refs. \cite{cafaro20,cafaroQR}%
, we suppose that the normalized output quantum state of a $\mathrm{su}\left(
2\text{; }%
\mathbb{C}
\right)  $ time-dependent Hamiltonian mimicking a continuous-time quantum
search algorithm can be described as
\begin{equation}
\left\vert \psi\left(  \theta\right)  \right\rangle \overset{\text{def}}%
{=}e^{i\varphi_{w}\left(  \theta\right)  }\sqrt{p_{w}\left(  \theta\right)
}\left\vert w\right\rangle +e^{i\varphi_{w_{\perp}}\left(  \theta\right)
}\sqrt{p_{w_{\perp}}\left(  \theta\right)  }\left\vert w_{\perp}\right\rangle
\text{,} \label{teta}%
\end{equation}
where the input is the normalized $N\overset{\text{def}}{=}2^{n}$-dimensional
$n$-qubit source state $\left\vert s\right\rangle \overset{\text{def}}%
{=}\left\vert \psi\left(  \theta_{0}\right)  \right\rangle $. Observe that
$\left\vert \psi\left(  \theta\right)  \right\rangle $ belongs to the
two-dimensional subspace of $\mathcal{H}_{2}^{n}$, the $n$-qubit complex
Hilbert space spanned by the set of orthonormal state vectors $\left\{
\left\vert w\right\rangle \text{, }\left\vert w_{\perp}\right\rangle \right\}
$ and containing $\left\vert s\right\rangle $. Furthermore, $\varphi
_{w}\left(  \theta\right)  $ and $\varphi_{w_{\perp}}\left(  \theta\right)  $
denote real quantum phases of the states $\left\vert w\right\rangle $ and
$\left\vert w_{\perp}\right\rangle $, respectively. Taking our source state
$\left\vert s\right\rangle $ to be identified with $\left\vert w_{\perp
}\right\rangle $, our analysis will focus on the space of probability
distributions $\left\{  p\left(  \theta\right)  \right\}  $ with $p\left(
\theta\right)  \overset{\text{def}}{=}\left(  p_{w}\left(  \theta\right)
\text{, }p_{w_{\perp}}\left(  \theta\right)  \right)  $ where $p_{w}\left(
\theta\right)  \overset{\text{def}}{=}\left\vert \left\langle w|\psi\left(
\theta\right)  \right\rangle \right\vert ^{2}$ and $p_{w_{\perp}}\left(
\theta\right)  \overset{\text{def}}{=}\left\vert \left\langle w_{\perp}%
|\psi\left(  \theta\right)  \right\rangle \right\vert ^{2}$ specify the
success and failure probabilities of the driving Hamiltonian, respectively.
For clarity, we underline that $\mathcal{X}\overset{\text{def}}{=}\left\{
x\right\}  =\left\{  w\text{, }w_{\perp}\right\}  $ forms here the space of
configuration variables with $p_{x}\left(  \theta\right)  $ being a
probability mass function since $\left\{  x\right\}  $ is a discrete set. In
particular, the space of probability distributions $\left\{  p\left(
\theta\right)  \right\}  $ is equipped with the natural Riemannian
distinguishability metric given by the Fisher information metric
$g_{\alpha\beta}\left(  \theta\right)  $. In the case of a discrete microspace
$\mathcal{X}$, $g_{\alpha\beta}\left(  \theta\right)  $ is defined as%
\begin{equation}
g_{\alpha\beta}\left(  \theta\right)  \overset{\text{def}}{=}\sum
_{x\in\mathcal{X}}p_{x}\left(  \theta\right)  \partial_{\alpha}\log\left[
p_{x}\left(  \theta\right)  \right]  \partial_{\beta}\log\left[  p_{x}\left(
\theta\right)  \right]  \text{.}%
\end{equation}
Furthermore, under suitably chosen working conditions \cite{caves94},
$g_{\alpha\beta}\left(  \theta\right)  $ can be taken to be proportional to
the Fubini-Study metric. Indeed, the Fubini-Study metric can be written as
$g_{\alpha\beta}^{\mathrm{FS}}\left(  \theta\right)  =(1/4)\left[
g_{\alpha\beta}\left(  \theta\right)  +4\sigma_{\alpha\beta}^{2}\left(
\theta\right)  \right]  \propto g_{\alpha\beta}\left(  \theta\right)  $ when
the variance of the phase changes $\sigma_{\alpha\beta}^{2}\left(
\theta\right)  $ is equal to zero. It happens that one can always set this
term equal to zero provided that one rephases in a favorable manner the basis
vectors used in the decomposition of $\left\vert \psi\left(  \theta\right)
\right\rangle $ as originally discussed in Ref. \cite{caves94}. We emphasize
that the output state $\left\vert \psi\left(  \theta\right)  \right\rangle $
is parametrized in terms of a single continuous real parameter $\theta$ that
emerges from the elapsed computing time $t$ of the algorithm (or,
equivalently, driving Hamiltonian). The parameter $\theta$, a statistical
version of $t$, plays the role of a statistical macrovariable employed to
distinguish neighboring quantum states $\left\vert \psi\left(  \theta\right)
\right\rangle $ and $\left\vert \psi\left(  \theta\right)  \right\rangle
+\left\vert d\psi\left(  \theta\right)  \right\rangle $ along a path through
the space of quantum mechanical pure states. It can be viewed as an
experimental parameter that can be determined by measurement of a conventional
observable that varies with time such as a time-dependent transverse magnetic
field intensity\textbf{. }Our main objective here is to calculate the IGC of
the optimum cooling paths, that is paths on the manifold of state space
parametrized by $\theta$ along which one drives the system while minimizing
the entropy production. Then, after evaluating the entropic efficiency of each
driving scheme being considered, we wish to find out whether or not there is
any link between this entropic efficiency and the IGC of the optimum cooling
paths generated by the driving schemes themselves.

In what follows, we describe how the normalized pure states $\left\{
\left\vert \psi\left(  \theta\right)  \right\rangle \right\}  $ that we
consider emerge as outputs of suitable $\mathrm{su}\left(  2\text{; }%
\mathbb{C}
\right)  $ time-dependent Hamiltonian evolutions that mimic quantum search
Hamiltonian motion.

\subsection{Quantum driving schemes}

We recall that $\mathrm{su}\left(  2\text{; }%
\mathbb{C}
\right)  $ is the Lie algebra of the special unitary group $\mathrm{SU}\left(
2\text{; }%
\mathbb{C}
\right)  $ and is generated by three traceless and anti-Hermitian generators
$\left\{  i\sigma_{x}\text{, }-i\sigma_{y}\text{, }i\sigma_{z}\right\}  $
where $\vec{\sigma}\overset{\text{def}}{=}\left(  \sigma_{x}\text{, }%
\sigma_{y}\text{, }\sigma_{z}\right)  $ is the Pauli vector operator
\cite{sakurai94}. We study quantum evolutions specified by means of
Hamiltonian operators $\mathcal{H}_{\mathrm{su}\left(  2\text{; }%
\mathbb{C}
\right)  }\left(  t\right)  $ defined as%
\begin{equation}
\mathcal{H}_{\mathrm{su}\left(  2\text{; }%
\mathbb{C}
\right)  }\left(  t\right)  \overset{\text{def}}{=}a\left(  t\right)  \left(
i\sigma_{x}\right)  +b\left(  t\right)  \left(  -i\sigma_{y}\right)  +c\left(
t\right)  \left(  \text{ }i\sigma_{z}\right)  \text{,}%
\end{equation}
with $a\left(  t\right)  $, $b\left(  t\right)  $, and $c\left(  t\right)  $
being time-dependent complex coefficients. Adopting the $\mathrm{su}\left(
2\text{; }%
\mathbb{C}
\right)  $-Hamiltonian models terminology, let us introduce the concepts of
complex transverse field and real longitudinal field, denoted as
$\omega\left(  t\right)  \overset{\text{def}}{=}\omega_{x}\left(  t\right)
-i\omega_{y}\left(  t\right)  =\omega_{\mathcal{H}}\left(  t\right)
e^{i\phi_{\omega}\left(  t\right)  }$ and $\Omega\left(  t\right)  $,
respectively. Obviously, $\omega_{\mathcal{H}}\left(  t\right)  $ represents
the modulus $\left\vert \omega\left(  t\right)  \right\vert $ of
$\omega\left(  t\right)  $. Then, setting $a\left(  t\right)  \overset
{\text{def}}{=}-i\omega_{x}\left(  t\right)  $, $b\left(  t\right)
\overset{\text{def}}{=}i\omega_{y}\left(  t\right)  $, and $c\left(  t\right)
\overset{\text{def}}{=}-i\Omega\left(  t\right)  $, the $\mathrm{su}\left(
2\text{; }%
\mathbb{C}
\right)  $-Hamiltonian becomes
\begin{equation}
\mathcal{H}_{\mathrm{su}\left(  2\text{; }%
\mathbb{C}
\right)  }\left(  t\right)  \overset{\text{def}}{=}\omega_{x}\left(  t\right)
\sigma_{x}+\omega_{y}\left(  t\right)  \sigma_{y}+\Omega\left(  t\right)
\sigma_{z}\text{.} \label{amo}%
\end{equation}
We assume that the transverse fields $\omega\left(  t\right)  $ lie in the
$xy$-plane while the longitudinal fields $\Omega\left(  t\right)  $ are
oriented\textbf{\ }along the $z$-axis. We observe that the $\mathrm{su}\left(
2\text{; }%
\mathbb{C}
\right)  $-Hamiltonian can be recast as $\mathcal{H}_{\mathrm{su}\left(
2\text{; }%
\mathbb{C}
\right)  }\left(  t\right)  \overset{\text{def}}{=}-\vec{\mu}\cdot\vec
{B}\left(  t\right)  $ when taking into consideration the evolution of a
spin-$1/2$ particle in an external time-dependent magnetic field $\vec
{B}\left(  t\right)  $. As usual, $\vec{\mu}\overset{\text{def}}{=}\left(
e\hslash/2mc\right)  \vec{\sigma}$ denotes the magnetic moment of the electron
with $\mu_{\text{Bohr}}\overset{\text{def}}{=}e\hslash/(2mc)$ being the
so-called Bohr magneton. The quantity $\left\vert e\right\vert $ denotes the
absolute value of the electric charge of an electron while $m$ is the mass of
an electron. Moreover, $\hslash$ and $c$ denote the reduced Planck constant
and the speed of light, respectively. To understand the relation between the
set of field intensities $\left\{  \omega_{\mathcal{H}}\left(  t\right)
\text{, }\Omega_{\mathcal{H}}\left(  t\right)  \right\}  $ and the magnetic
field $\vec{B}\left(  t\right)  $, we decompose $\vec{B}\left(  t\right)  $ as
$\vec{B}\left(  t\right)  \overset{\text{def}}{=}\vec{B}_{\perp}\left(
t\right)  +\vec{B}_{\parallel}\left(  t\right)  $, with $\vec{B}_{\perp
}\left(  t\right)  \overset{\text{def}}{=}B_{x}\left(  t\right)  \hat{x}%
+B_{y}\left(  t\right)  \hat{y}$ and $\vec{B}_{\parallel}\left(  t\right)
\overset{\text{def}}{=}B_{z}\left(  t\right)  \hat{z}$. Then, it follows that
$B_{\perp}\left(  t\right)  \propto\omega_{\mathcal{H}}\left(  t\right)
\overset{\text{def}}{=}\left\vert \omega\left(  t\right)  \right\vert $ and
$B_{\parallel}\left(  t\right)  \propto$ $\Omega_{\mathcal{H}}\left(
t\right)  \overset{\text{def}}{=}\left\vert \Omega\left(  t\right)
\right\vert $. More specifically, the exact relation in terms of field
components between $\left\{  B_{x}\left(  t\right)  \text{, }B_{y}\left(
t\right)  \text{, }B_{z}\left(  t\right)  \right\}  $ and $\left\{  \omega
_{x}\left(  t\right)  \text{, }\omega_{y}\left(  t\right)  \text{, }%
\Omega\left(  t\right)  \right\}  $ is expressed by the equalities
\begin{equation}
B_{x}\left(  t\right)  =-\frac{2mc}{e\hslash}\omega_{x}\left(  t\right)
\text{, }B_{y}\left(  t\right)  =-\frac{2mc}{e\hslash}\omega_{y}\left(
t\right)  \text{, and }B_{z}\left(  t\right)  =-\frac{2mc}{e\hslash}%
\Omega\left(  t\right)  \text{.}%
\end{equation}
Furthermore, in terms of field intensities $B_{\perp}\left(  t\right)  $ and
$B_{\parallel}\left(  t\right)  $, one obtains
\begin{equation}
B_{\perp}\left(  t\right)  =\frac{2mc}{\left\vert e\right\vert \hslash}%
\omega_{\mathcal{H}}\left(  t\right)  \text{, and }B_{\parallel}\left(
t\right)  =\frac{2mc}{\left\vert e\right\vert \hslash}\Omega_{\mathcal{H}%
}\left(  t\right)  \text{.}%
\end{equation}
Despite its apparent simplicity, it is a highly challenging matter studying
the evolution of an electron specified by the Hamiltonian $\mathcal{H}%
_{\mathrm{su}\left(  2\text{; }%
\mathbb{C}
\right)  }\left(  t\right)  $ by means of exact analytical expressions of
complex probability amplitudes and real transition probabilities from an
initial source state to a final target state. The quantum mechanical time
propagator $\mathcal{U}\left(  t\right)  $,%
\begin{equation}
\mathcal{U}\left(  t\right)  \overset{\text{def}}{=}\left(
\begin{array}
[c]{cc}%
\alpha\left(  t\right)  & \beta\left(  t\right) \\
-\beta^{\ast}\left(  t\right)  & \alpha^{\ast}\left(  t\right)
\end{array}
\right)  \text{,}%
\end{equation}
with $i\hslash\mathcal{\dot{U}}\left(  t\right)  =\mathcal{H}_{\mathrm{su}%
\left(  2\text{; }%
\mathbb{C}
\right)  }\mathcal{U}\left(  t\right)  $ and $\mathcal{\dot{U}}\overset
{\text{def}}{=}\partial_{t}\mathcal{U}$, is unitary and demands that the
probability amplitudes $\alpha\left(  t\right)  $ and $\beta\left(  t\right)
$ must satisfy the normalization condition $\left\vert \alpha\left(  t\right)
\right\vert ^{2}+\left\vert \beta\left(  t\right)  \right\vert ^{2}=1$. Then,
\textbf{ }being $\left\{  \left\vert w\right\rangle \text{, }\left\vert
w_{\perp}\right\rangle \right\}  $ a set of orthonormal state vectors that
span the two-dimensional search\textbf{ }space\textbf{ }of the\textbf{
}$\mathcal{H}_{2}^{n}$, the time evolution of a source state $\left\vert
s\right\rangle \overset{\text{def}}{=}x\left\vert w\right\rangle
+\sqrt{1-x^{2}}\left\vert w_{\perp}\right\rangle $ with $x\overset{\text{def}%
}{=}\left\langle w|s\right\rangle $ can be described by the mapping,%
\begin{equation}
\left(  x\text{, }\sqrt{1-x^{2}}\right)  \overset{\mathcal{U}\left(  t\right)
}{\rightarrow}\left(  \alpha\left(  t\right)  x+\beta\left(  t\right)
\sqrt{1-x^{2}}\text{, }-\beta^{\ast}\left(  t\right)  x+\alpha^{\ast}\left(
t\right)  \sqrt{1-x^{2}}\right)  \text{.}%
\end{equation}
Thus, the probability $\mathcal{P}_{\left\vert s\right\rangle \rightarrow
\left\vert w\right\rangle }\left(  t\right)  $ that under $\mathcal{U}\left(
t\right)  $ the source state $\left\vert s\right\rangle $ transitions into the
target state $\left\vert w\right\rangle $ becomes,%
\begin{equation}
\mathcal{P}_{\left\vert s\right\rangle \rightarrow\left\vert w\right\rangle
}\left(  t\right)  \overset{\text{def}}{=}\left\vert \left\langle
w|\mathcal{U}\left(  t\right)  |s\right\rangle \right\vert ^{2}=\left\vert
\alpha\left(  t\right)  \right\vert ^{2}x^{2}+\left\vert \beta\left(
t\right)  \right\vert ^{2}\left(  1-x^{2}\right)  +\left[  \alpha\left(
t\right)  \beta^{\ast}\left(  t\right)  +\alpha^{\ast}\left(  t\right)
\beta\left(  t\right)  \right]  x\sqrt{1-x^{2}}\text{.} \label{good}%
\end{equation}
As evident from Eq. (\ref{good}), it is necessary to possess the exact
analytical expression of the evolution operator $\mathcal{U}\left(  t\right)
$ in terms of the complex probability amplitudes $\alpha\left(  t\right)  $
and $\beta\left(  t\right)  $ to calculate the exact analytical expression of
$\mathcal{P}_{\left\vert s\right\rangle \rightarrow\left\vert w\right\rangle
}\left(  t\right)  $. For completeness, a general parametrization of
$\alpha\left(  t\right)  $ and $\beta\left(  t\right)  $ is given in
Appendix\ D. Inspired by our results reported in Ref. \cite{alsing19} and,
above all, making use of the findings in Refs. \cite{messina14,grimaudo18}, we
focus our attention on four distinct quantum mechanical driving scenarios
where $\mathcal{P}_{\left\vert w_{\perp}\right\rangle \rightarrow\left\vert
w\right\rangle }\left(  t\right)  $ can be analytically expressed. The states
$\left\vert w\right\rangle $ and $\left\vert w_{\perp}\right\rangle $ with
$\left\langle w_{\perp}|w\right\rangle =\delta_{w_{\perp}\text{, }w}$ are
chosen so that $\sigma_{z}\left\vert w\right\rangle =+\left\vert
w\right\rangle $ and $\sigma_{z}\left\vert w_{\perp}\right\rangle =-\left\vert
w_{\perp}\right\rangle $. The quantity that specifies the four scenarios is
the modulus $\left\vert \omega\left(  t\right)  \right\vert $ of the complex
transverse field $\omega\left(  t\right)  $, $\omega_{\mathcal{H}}\left(
t\right)  \overset{\text{def}}{=}\left\vert \omega\left(  t\right)
\right\vert \propto B_{\bot}\left(  t\right)  $. However, for experimental
convenience, we assume that $\dot{\phi}_{\omega}\left(  t\right)  =\omega_{0}$
and $\Omega\left(  t\right)  =-\left(  \hslash/2\right)  \omega_{0}$ with
$\omega_{0}$ a negative constant in all four scenarios. More general temporal
behaviors $\dot{\phi}_{\omega}\left(  t\right)  $ and $\Omega\left(  t\right)
$ can be chosen provided that the so-called generalized Rabi condition
$\dot{\phi}_{\omega}\left(  t\right)  +(2/\hslash)\Omega\left(  t\right)  =$
$0$ is satisfied as pointed out in Ref. \cite{messina14,grimaudo18}. The first
case specifies the original Rabi scenario where we assume a constant field
intensity $\omega_{\mathcal{H}}^{\left(  1\right)  }\left(  t\right)
\overset{\text{def}}{=}\Gamma$ with $\mathcal{P}_{\left\vert w_{\perp
}\right\rangle \rightarrow\left\vert w\right\rangle }^{\left(  1\right)
}\left(  t\right)  =\sin^{2}\left[  \left(  \Gamma/\hslash\right)  t\right]
$. The remaining three cases are generalized Rabi scenarios with field
intensity assumed to be exhibiting oscillatory, power law decay, and
exponential law decay behaviors. In summary, we have%
\begin{equation}
\omega_{\mathcal{H}}^{\left(  1\right)  }\left(  t\right)  \overset
{\text{def}}{=}\Gamma\text{, }\omega_{\mathcal{H}}^{\left(  2\right)  }\left(
t\right)  \overset{\text{def}}{=}\Gamma\cos\left(  \lambda t\right)  \text{,
}\omega_{\mathcal{H}}^{\left(  3\right)  }\left(  t\right)  \overset
{\text{def}}{=}\Gamma/\left(  1+\lambda t\right)  ^{2}\text{, and }%
\omega_{\mathcal{H}}^{\left(  4\right)  }\left(  t\right)  \overset
{\text{def}}{=}\Gamma e^{-\lambda t}\text{.} \label{FI}%
\end{equation}
Note that $\omega_{\mathcal{H}}^{\left(  2\right)  }\left(  t\right)  \geq0$
for $0\leq t\leq\left(  \pi/2\right)  \lambda^{-1}$. In all four cases, it
happens that $\mathcal{P}_{\left\vert w_{\perp}\right\rangle \rightarrow
\left\vert w\right\rangle }^{\left(  j\right)  }\left(  t\right)  $ with
$1\leq j\leq4$ is given by \cite{grimaudo18}%
\begin{equation}
\mathcal{P}_{\left\vert w_{\perp}\right\rangle \rightarrow\left\vert
w\right\rangle }^{\left(  j\right)  }\left(  t\right)  =\sin^{2}\left[
\int_{0}^{t}\frac{\omega_{\mathcal{H}}^{\left(  j\right)  }\left(  t^{\prime
}\right)  }{\hslash}dt^{\prime}\right]  \text{.} \label{tp2}%
\end{equation}
Interestingly, since the resonance condition is satisfied, $\mathcal{P}%
_{\left\vert w_{\perp}\right\rangle \rightarrow\left\vert w\right\rangle
}^{\left(  j\right)  }\left(  t\right)  $ in Eq. (\ref{tp2}) depends only on
the integral of the transverse field intensity $\omega_{\mathcal{H}}\left(
t\right)  $. The transition probabilities $\mathcal{P}_{\left\vert w_{\perp
}\right\rangle \rightarrow\left\vert w\right\rangle }^{\left(  k\right)
}\left(  t\right)  $ in Eq. (\ref{tp2}) are the key ingredients that we
exploit to provide an expression of the parametrized output quantum states
$\left\vert \psi\left(  \theta\right)  \right\rangle $.

Having introduced the IGC in Eq. (\ref{IGC}), the entropic efficiency in Eq.
(\ref{efficiency}), and our chosen quantum mechanical driving schemes, we are
ready to apply our proposed theoretical analysis.

\section{Applications}

In this section, we apply our theoretical construct to four distinct quantum
mechanical driving scenarios.

To apply our scheme, we need to find the optimum cooling (probability) paths
before evaluating the information geometric complexity $\mathcal{C}%
_{\mathcal{M}_{s}}$ in Eq. (\ref{IGC}), the entropic speed $v_{\mathrm{E}}$ in
Eq. (\ref{vthermo}), the rate of entropy production $r_{\mathrm{E}}$ in Eq.
(\ref{EPR}), and the entropic efficiency $\eta_{\mathrm{E}}$ in\ Eq.
(\ref{efficiency}) along these geodesic trajectories. To find these paths
$\gamma_{\theta}:\theta\mapsto p\left(  \theta\right)  $ with $\theta
=\theta\left(  \xi\right)  $ and $\xi$ being an affine parameter, we proceed
as follows. For each Schr\"{o}dinger evolution characterized by a specific
expression of $\omega_{\mathcal{H}}\left(  t\right)  $ (that is, the modulus
of the complex transverse field $\omega\left(  t\right)  $ that is
proportional to $B_{\bot}\left(  t\right)  $), we arrive at the regular
probability paths $\left\{  p\left(  \theta\right)  \right\}  $ with $p\left(
\theta\right)  \overset{\text{def}}{=}\left(  p_{w}\left(  \theta\right)
\text{, }p_{w_{\perp}}\left(  \theta\right)  \right)  $ as prescribed in the
previous section. Then, having $\left\{  p\left(  \theta\right)  \right\}  $,
we calculate the Fisher information $g\left(  \theta\right)  =E_{\theta
}\left[  \left\{  \partial_{\theta}\log\left[  p_{x}\left(  \theta\right)
\right]  \right\}  ^{2}\right]  $ with $E_{\theta}\left[  \mathrm{V}\right]  $
denoting the expected value of the random variable $\mathrm{V}$ with respect
to the probability mass function $p_{x}\left(  \theta\right)  $ along these
probability paths. The Fisher information enters the geodesic equation for
$\theta=\theta\left(  \xi\right)  $. Finally, upon integrating the geodesic
equation, we find the most favorable time parametrizations of $\gamma_{\theta
}$ and, consequently, the optimum cooling paths $\left\{  p_{\mathrm{optimum}%
}\left(  \theta\right)  \right\}  $.

Before starting our geodesic analysis, we recall for completeness that the
trajectories connecting two quantum states $\left\vert A\right\rangle $ and
$\left\vert B\right\rangle $ generated by an optimal-speed unitary evolution
$U$ can be regarded as geodesic curves on the Bloch sphere. From a geometric
standpoint, these unitary operators $\left\{  U\right\}  $ can be described by
means of rotations of the Bloch sphere around the axis that is orthogonal to
the hemispherical plane containing the origin along with $\left\vert
A\right\rangle $ and $\left\vert B\right\rangle $ \cite{brody03}. In our
paper, instead, optimality means minimum entropy production and not
time-optimality. In addition, minimum entropy production probability paths are
geodesic paths on the parametric manifold with elements specified by the
parameter $\theta$ and not on the Bloch sphere of pure quantum states.

\begin{figure}[t]
\centering
\includegraphics[width=1\textwidth] {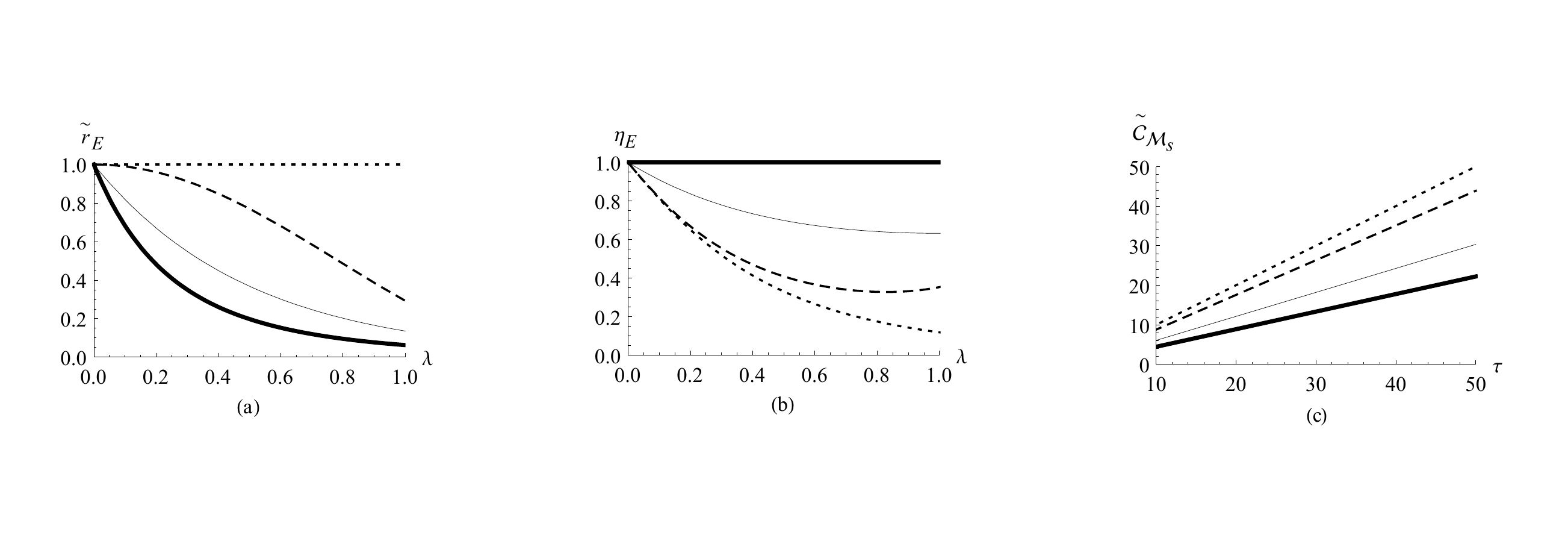}\caption{We plot in (a) the rescaled
entropy production rate $\tilde{r}_{\mathrm{E}}$ versus $\lambda$, the
parameter that characterizes the field intensity. In (b), we depict the
behavior of the entropic efficiency $\eta_{\mathrm{E}}$ versus $\lambda$. In
(c), we plot $\mathcal{\tilde{C}}_{\mathcal{M}_{s}}$ versus $\tau$ with
$\mathcal{\tilde{C}}_{\mathcal{M}_{s}}$ being the rescaled version of the
information geometric complexity $\mathcal{C}_{\mathcal{M}_{s}}$ in its
long-time limit. In (a), (b), and (c), the dotted, dashed, thin solid, and
thick solid lines correspond to the constant, oscillatory, exponential law
decay, and power law decay field intensity behaviors, respectively. Finally,
we set $\theta_{0}=1$ in all plots and $\lambda=1/2$ in (c).}%
\end{figure}

\subsubsection{Constant $\omega_{\mathcal{H}}$}

The first driving scheme that we consider is characterized by a constant
$\omega_{\mathcal{H}}^{\left(  1\right)  }\left(  t\right)  =\Gamma$. In this
case, the space of probability distributions $\left\{  p\left(  \theta\right)
\right\}  $ with $p\left(  \theta\right)  \overset{\text{def}}{=}\left(
p_{w}\left(  \theta\right)  \text{, }p_{w_{\perp}}\left(  \theta\right)
\right)  $ is specified by the success and failure probabilities%
\begin{equation}
p_{w}\left(  \theta\right)  \overset{\text{def}}{=}\sin^{2}\left(
\frac{\Gamma}{\hslash}\theta\right)  \text{, and }p_{w_{\perp}}\left(
\theta\right)  \overset{\text{def}}{=}\cos^{2}\left(  \frac{\Gamma}{\hslash
}\theta\right)  \text{,} \label{1a}%
\end{equation}
respectively. The probabilities in Eq. (\ref{1a}) present a periodic
oscillatory behavior with period $T\overset{\text{def}}{=}\left(  \pi
\hslash\right)  /\Gamma$ while the Fisher information $g\left(  \theta\right)
$ assumes the constant value $g_{0}$ $\overset{\text{def}}{=}\left(
2\Gamma/\hslash\right)  ^{2}$. Finally, the geodesic equations yielding the
most favorable time parametrizations of $\gamma_{\theta}$ becomes
\begin{equation}
\frac{d^{2}\theta}{d\xi^{2}}+\frac{1}{2g}\frac{dg}{d\theta}\left(
\frac{d\theta}{d\xi}\right)  ^{2}=0\text{.} \label{odeti}%
\end{equation}
Given that $g\left(  \theta\right)  =g_{0}$ together with supposing non
vanishing positive initial conditions $\theta\left(  \xi_{0}\right)
=\theta_{0}$ and $\dot{\theta}\left(  \xi_{0}\right)  =\dot{\theta}_{0}$,
integration of the geodesic equation leads to the following optimum paths,%
\begin{equation}
\theta\left(  \xi\right)  =\theta_{0}+\dot{\theta}_{0}\left(  \xi-\xi
_{0}\right)  \text{.} \label{geo1a}%
\end{equation}
From the expression of the optimum paths in\ Eq. (\ref{geo1a}), we can
evaluate $\mathcal{C}_{\mathcal{M}_{s}}$ in Eq. (\ref{IGC}), $v_{\mathrm{E}}$
in Eq. (\ref{vthermo}), $r_{\mathrm{E}}$ in Eq. (\ref{EPR}), and
$\eta_{\mathrm{E}}$ in\ Eq. (\ref{efficiency}). Specifically, we get%
\begin{equation}
\mathcal{C}_{\mathcal{M}_{s}}\left(  \tau\right)  =\frac{\Gamma}{\hslash
}\left(  \tau+\tau_{0}-2\xi_{0}\right)  \dot{\theta}_{0}\text{, }%
v_{\mathrm{E}}\left(  \Gamma\right)  =\frac{2\Gamma}{\hslash}\dot{\theta}%
_{0}\text{, and }r_{\mathrm{E}}\left(  \Gamma\right)  =\left(  \frac{2\Gamma
}{\hslash}\right)  ^{2}\dot{\theta}_{0}^{2}\text{.} \label{speed1}%
\end{equation}
From Eq. (\ref{speed1}), we observe that $v_{\mathrm{E}}\left(  \Gamma\right)
\propto\Gamma$, $r_{\mathrm{E}}\left(  \Gamma\right)  \propto\Gamma^{2}$, and
$\mathcal{C}_{\mathcal{M}_{s}}\left(  \tau\right)  $ grows linearly in time
with $d\mathcal{C}_{\mathcal{M}_{s}}/d\tau\propto v_{\mathrm{E}}%
=r_{\mathrm{E}}^{1/2}$. It is transparent from Eq. (\ref{speed1}) that
$\omega_{\mathcal{H}}^{\left(  1\right)  }\left(  t\right)  =\Gamma$, the
modulus of the complex transverse field that specifies the $\mathrm{su}\left(
2\text{; }%
\mathbb{C}
\right)  $ driving Hamiltonian, is the parameter to be tuned in order to find
a suitable tradeoff between speed and efficiency (or, speed and information
geometric complexity) within our analysis of quantum mechanical evolutions.
For clarity, we emphasize that the expression of $r_{\mathrm{E}}$ in Eq.
(\ref{speed1}) can be obtained either from Eq. (\ref{EPR}) or Eq. (\ref{er3}).
In particular, we point out that the rate of entropy production inherits the
typical initial-state dependence of the entropy production \cite{gu21} as
evident from its expression in Eq. (\ref{speed1}). These last two
clarifications apply to all quantum driving scenarios that we study here.
Moreover, we remark that the linear growth with respect to the temporal
variable $\tau$ of the IGC is not completely unexpected. Indeed, in all cases
being considered here, there is only one control parameter $\theta$ and,
roughly speaking, the explored parametric volumes reduce to explored lengths.
Finally, since the motion is geodesic, the covariant acceleration vanishes and
the evolution of the control parameter occurs with constant entropic speed.
For this reason, to compare the various driving schemes using the IGC, the
quantity that gains more relevance is the rate of change\textbf{
}$d\mathcal{C}_{\mathcal{M}_{s}}/d\tau$\textbf{ }of the IGCs with respect to
$\tau$.

\subsubsection{Oscillating $\omega_{\mathcal{H}}$}

The second driving scheme that we study is specified by $\omega_{\mathcal{H}%
}^{\left(  2\right)  }\left(  t\right)  =\Gamma\cos\left(  \lambda t\right)  $
with $\lambda\in%
\mathbb{R}
_{+}$ being a frequency parameter. In this case, the space of probability
distributions $\left\{  p\left(  \theta\right)  \right\}  $ with $p\left(
\theta\right)  \overset{\text{def}}{=}\left(  p_{w}\left(  \theta\right)
\text{, }p_{w_{\perp}}\left(  \theta\right)  \right)  $ is given by%
\begin{equation}
p_{w}\left(  \theta\right)  \overset{\text{def}}{=}\sin^{2}\left[
\frac{\Gamma}{\hslash\lambda}\sin\left(  \lambda\theta\right)  \right]
\text{, and }p_{w_{\perp}}\left(  \theta\right)  \overset{\text{def}}{=}%
\cos^{2}\left[  \frac{\Gamma}{\hslash\lambda}\sin\left(  \lambda\theta\right)
\right]  \text{,} \label{5b}%
\end{equation}
respectively. The probabilities $p_{w}\left(  \theta\right)  $ and
$p_{w_{\perp}}\left(  \theta\right)  $ exhibit a periodic oscillatory behavior
with period given by $T\overset{\text{def}}{=}\pi/\lambda$. Furthermore, since
$p_{w}\left(  \theta\right)  $ reaches its maximum value $\sin^{2}\left[
\Gamma/\left(  \hslash\lambda\right)  \right]  $ at $t^{\ast}\overset
{\text{def}}{=}\pi/\left(  2\lambda\right)  $, we must impose the constraint
$\Gamma/\lambda=h/4$ in order for $p_{w}\left(  \theta\right)  $ to reach one
as its maximum value. From Eq. (\ref{5b}), we get $g\left(  \theta\right)
=\left(  2\Gamma/\hslash\right)  ^{2}\cos^{2}\left(  \lambda\theta\right)  $
while the geodesic equation becomes%
\begin{equation}
\frac{d^{2}\theta}{d\xi^{2}}-\lambda\tan\left(  \lambda\theta\right)  \left(
\frac{d\theta}{d\xi}\right)  ^{2}=0\text{.}%
\end{equation}
Remaining in the working assumptions of nonvanishing positive initial
conditions $\theta\left(  \xi_{0}\right)  =\theta_{0}$ and $\dot{\theta
}\left(  \xi_{0}\right)  =\dot{\theta}_{0}$, integration of the geodesic
equation leads to optimum paths $\theta\left(  \xi\right)  $\textbf{ }of the
form,%
\begin{equation}
\theta\left(  \xi\right)  =\frac{1}{\lambda}\sin^{-1}\left[  \lambda
\cos\left(  \lambda\theta_{0}\right)  \left(  \xi-\xi_{0}\right)  \dot{\theta
}_{0}+\sin\left(  \lambda\theta_{0}\right)  \right]  \text{.} \label{geo2}%
\end{equation}
As pointed out earlier, from the optimum paths in\ Eq. (\ref{geo2}), we can
evaluate $\mathcal{C}_{\mathcal{M}_{s}}$ in Eq. (\ref{IGC}), $v_{\mathrm{E}}$
in Eq. (\ref{vthermo}), $r_{\mathrm{E}}$ in Eq. (\ref{EPR}), and
$\eta_{\mathrm{E}}$ in\ Eq. (\ref{efficiency}). We obtain,%
\begin{equation}
\mathcal{C}_{\mathcal{M}_{s}}\left(  \tau\right)  =\frac{\Gamma}{\hslash
}\left(  \tau+\tau_{0}-2\xi_{0}\right)  \dot{\theta}_{0}\left\vert \cos\left(
\lambda\theta_{0}\right)  \right\vert \text{, }v_{\mathrm{E}}\left(
\Gamma\right)  =\frac{2\Gamma}{\hslash}\left\vert \cos\left(  \lambda
\theta_{0}\right)  \right\vert \dot{\theta}_{0}\text{, and }r_{\mathrm{E}%
}\left(  \Gamma\right)  =\left(  \frac{2\Gamma}{\hslash}\right)  ^{2}\cos
^{2}\left(  \lambda\theta_{0}\right)  \dot{\theta}_{0}^{2}\text{,}
\label{speed1b}%
\end{equation}
where $\lambda=\lambda\left(  \Gamma\right)  \overset{\text{def}}{=}\left(
4\Gamma\right)  /h$. From Eqs. (\ref{speed1b}) and (\ref{speed1}), we notice
that the IGC keeps growing linearly in time with $d\mathcal{C}_{\mathcal{M}%
_{s}}/d\tau\propto v_{\mathrm{E}}=r_{\mathrm{E}}^{1/2}$. The geodesic motion,
however, yields cooler optimum paths that are explored with a smaller entropic speed.

\subsubsection{Power law decay of $\omega_{\mathcal{H}}$}

The third driving scheme is characterized by $\omega_{\mathcal{H}}^{\left(
3\right)  }\left(  t\right)  =\Gamma/\left(  1+\lambda t\right)  ^{2}$. In
this case, the space of probability distributions $\left\{  p\left(
\theta\right)  \right\}  $ is given by $p\left(  \theta\right)  \overset
{\text{def}}{=}\left(  p_{w}\left(  \theta\right)  \text{, }p_{w_{\perp}%
}\left(  \theta\right)  \right)  $ with
\begin{equation}
p_{w}\left(  \theta\right)  \overset{\text{def}}{=}\sin^{2}\left[
\frac{\Gamma}{\hslash\lambda}\left(  1-\frac{1}{1+\lambda\theta}\right)
\right]  \text{, and }p_{w_{\perp}}\left(  \theta\right)  \overset{\text{def}%
}{=}\cos^{2}\left[  \frac{\Gamma}{\hslash\lambda}\left(  1-\frac{1}%
{1+\lambda\theta}\right)  \right]  \text{,} \label{4b}%
\end{equation}
respectively. Provided that $\Gamma/\lambda=h/4$, $p_{w}\left(  \theta\right)
$ in Eq. (\ref{4b}) exhibits an asymptotic monotonic convergence to one.
Moreover, the Fisher information is given by $g\left(  \theta\right)  =\left(
2\Gamma/\hslash\right)  ^{2}\left(  1+\lambda\theta\right)  ^{-4}$ while the
geodesic equation is%
\begin{equation}
\frac{d^{2}\theta}{d\xi^{2}}-\frac{2\lambda}{1+\lambda\theta}\left(
\frac{d\theta}{d\xi}\right)  ^{2}=0\text{.}%
\end{equation}
As previously mentioned, we keep assuming nonvanishing positive initial
conditions $\theta\left(  \xi_{0}\right)  =\theta_{0}$ and $\dot{\theta
}\left(  \xi_{0}\right)  =\dot{\theta}_{0}$. Then, integrating the geodesic
equation, we obtain the optimum paths given by%
\begin{equation}
\theta\left(  \xi\right)  =\frac{\left(  1+\lambda\theta_{0}\right)
^{2}+\lambda\dot{\theta}_{0}\left[  \left(  \xi-\xi_{0}\right)  -\frac
{1+\lambda\theta_{0}}{\lambda\dot{\theta}_{0}}\right]  }{\lambda^{2}%
\dot{\theta}_{0}\left[  \frac{1+\lambda\theta_{0}}{\lambda\dot{\theta}_{0}%
}-\left(  \xi-\xi_{0}\right)  \right]  }\text{.} \label{geo3}%
\end{equation}
From the optimum paths in\ Eq. (\ref{geo3}), we compute $\mathcal{C}%
_{\mathcal{M}_{s}}$ in Eq. (\ref{IGC}), $v_{\mathrm{E}}$ in Eq. (\ref{vthermo}%
), $r_{\mathrm{E}}$ in Eq. (\ref{EPR}), and $\eta_{\mathrm{E}}$ in\ Eq.
(\ref{efficiency}). In particular, we get%
\begin{equation}
\mathcal{C}_{\mathcal{M}_{s}}\left(  \tau\right)  =\frac{\Gamma}{\hslash
}\left(  \tau+\tau_{0}-2\xi_{0}\right)  \dot{\theta}_{0}\frac{1}{\left[
1+\lambda\left(  \Gamma\right)  \theta_{0}\right]  ^{2}}\text{, }%
v_{\mathrm{E}}\left(  \Gamma\right)  =\frac{2\Gamma}{\hslash}\frac{1}{\left[
1+\lambda\left(  \Gamma\right)  \theta_{0}\right]  ^{2}}\dot{\theta}%
_{0}\text{, and }r_{\mathrm{E}}\left(  \Gamma\right)  =\left(  \frac{2\Gamma
}{\hslash}\right)  ^{2}\frac{1}{\left[  1+\lambda\left(  \Gamma\right)
\theta_{0}\right]  ^{4}}\dot{\theta}_{0}^{2}\text{,} \label{speed3}%
\end{equation}
with $\lambda\left(  \Gamma\right)  \overset{\text{def}}{=}\left(
4\Gamma\right)  /h$. Analogously to the first and second scenarios, the motion
on the manifold associated with the third scenario proceeds at constant
entropic speed $v_{\mathrm{E}}$ and, thus, exhibits minimum entropy
production. The IGC keeps growing linearly in time with $d\mathcal{C}%
_{\mathcal{M}_{s}}/d\tau\propto v_{\mathrm{E}}=r_{\mathrm{E}}^{1/2}$. In
particular, this third scenario is characterized by a geodesic motion that
gives rise to optimum paths that are cooler than those corresponding to the
second scenario.

\begin{figure}[t]
\centering
\includegraphics[width=1\textwidth] {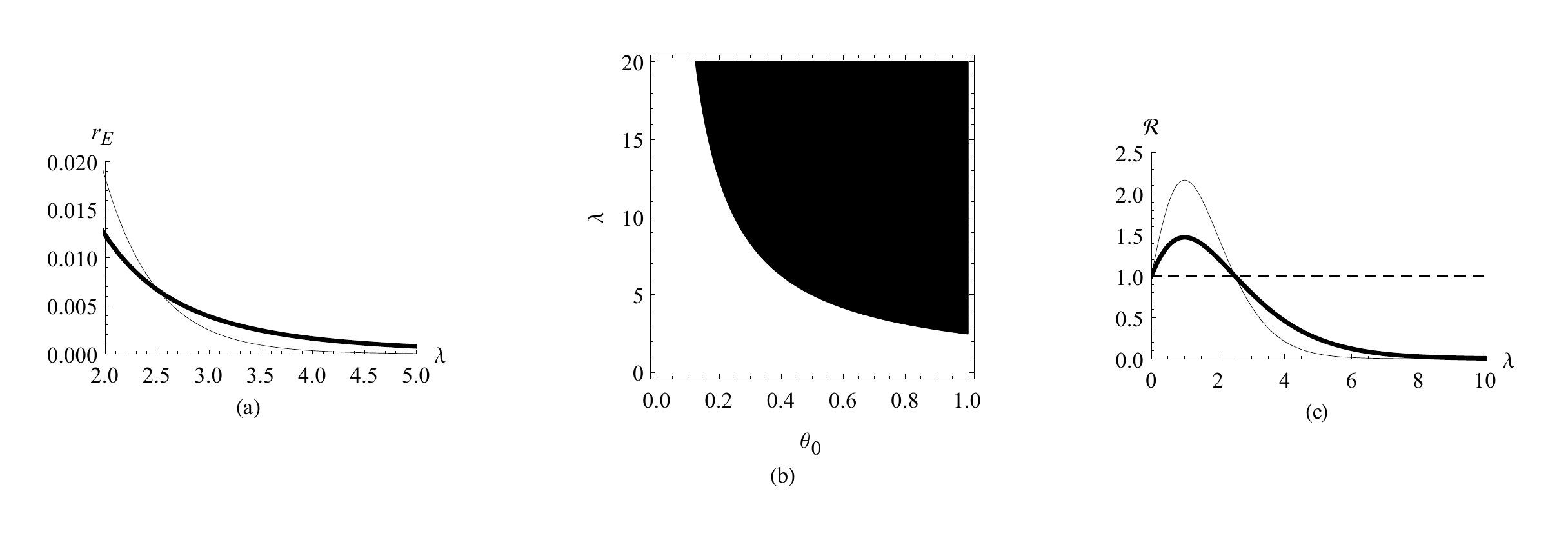}\caption{In (a), we plot the entropy
production rate $r_{\mathrm{E}}$ versus $\lambda$ and set $\theta_{0}=1$. The
thin solid and thick solid lines represent the exponential law and the power
law decay scenarios, respectively. The intersection between the two lines
occurs at $\lambda\simeq2.51$. In (b), we plot the parametric region
$\mathcal{D}\left(  \theta_{0}\text{, }\lambda\right)  $ where $r_{\mathrm{E}%
}^{\left(  \text{\textrm{exponential}}\right)  }\left(  \theta_{0}\text{,
}\lambda\right)  \leq r_{\mathrm{E}}^{\left(  \text{\textrm{power-law}%
}\right)  }\left(  \theta_{0}\text{, }\lambda\right)  $ (black region).
Finally, in (c) we plot the ratios \textrm{R}$_{\mathcal{C}_{\mathcal{M}_{s}}%
}\overset{\text{def}}{=}\mathcal{C}_{\mathcal{M}_{s}}^{\left(
\text{\textrm{exponential}}\right)  }/\mathcal{C}_{\mathcal{M}_{s}}^{\left(
\text{\textrm{power-law}}\right)  }$ (thick solid line) and \textrm{R}%
$_{r_{\mathrm{E}}}\overset{\text{def}}{=}r_{\mathrm{E}}^{\left(
\text{\textrm{exponential}}\right)  }/r_{\mathrm{E}}^{\left(
\text{\textrm{power-law}}\right)  }$ (thin solid line) versus $\lambda$ with
$\theta_{0}$ set equal to one. We note that the exponential law decay scheme
outperforms the power law decay scheme in terms of both entropy production
rate and information geometric complexity in the limit of sufficiently large
values of $\lambda$.}%
\end{figure}

\subsubsection{Exponential decay of $\omega_{\mathcal{H}}$}

The fourth driving scheme is characterized by $\omega_{\mathcal{H}}^{\left(
4\right)  }\left(  t\right)  =\Gamma e^{-\lambda t}$. In this case, the space
of probability distributions $\left\{  p\left(  \theta\right)  \right\}  $ is
given by $p\left(  \theta\right)  \overset{\text{def}}{=}\left(  p_{w}\left(
\theta\right)  \text{, }p_{w_{\perp}}\left(  \theta\right)  \right)  $ where%
\begin{equation}
p_{w}\left(  \theta\right)  \overset{\text{def}}{=}\sin^{2}\left[
\frac{\Gamma}{\hslash\lambda}\left(  1-e^{-\lambda\theta}\right)  \right]
\text{, and }p_{w_{\perp}}\left(  \theta\right)  \overset{\text{def}}{=}%
\cos^{2}\left[  \frac{\Gamma}{\hslash\lambda}\left(  1-e^{-\lambda\theta
}\right)  \right]  \text{,} \label{3aa}%
\end{equation}
respectively. We note that as long as $\Gamma/\lambda=h/4$, the probability
$p_{w}\left(  \theta\right)  $ in Eq. (\ref{3aa}) presents an asymptotic
monotonic convergence to one. Employing Eq. (\ref{3aa}), the Fisher
information becomes $g\left(  \theta\right)  =\left(  2\Gamma/\hslash\right)
^{2}e^{-2\lambda\theta}$ and the geodesic equation is%
\begin{equation}
\frac{d^{2}\theta}{d\xi^{2}}-\lambda\left(  \frac{d\theta}{d\xi}\right)
^{2}=0\text{.}%
\end{equation}
Integrating the geodesic equation and assuming a set of nonvanishing positive
initial conditions $\theta\left(  \xi_{0}\right)  =\theta_{0}$ and
$\dot{\theta}\left(  \xi_{0}\right)  =\dot{\theta}_{0}$, the optimum paths
become%
\begin{equation}
\theta\left(  \xi\right)  =\theta_{0}-\frac{1}{\lambda}\log\left[
1-\lambda\dot{\theta}_{0}\left(  \xi-\xi_{0}\right)  \right]  \text{.}
\label{geo4}%
\end{equation}
From the optimum paths in\ Eq. (\ref{geo4}), we calculate $\mathcal{C}%
_{\mathcal{M}_{s}}$ in Eq. (\ref{IGC}), $v_{\mathrm{E}}$ in Eq. (\ref{vthermo}%
), $r_{\mathrm{E}}$ in Eq. (\ref{EPR}), and $\eta_{\mathrm{E}}$ in\ Eq.
(\ref{efficiency}). In particular, we obtain%
\begin{equation}
\mathcal{C}_{\mathcal{M}_{s}}\left(  \tau\right)  =\frac{\Gamma}{\hslash
}\left(  \tau+\tau_{0}-2\xi_{0}\right)  \dot{\theta}_{0}e^{-\lambda\theta_{0}%
}\text{, }v_{\mathrm{E}}\left(  \Gamma\right)  =\frac{2\Gamma}{\hslash
}e^{-\lambda\left(  \Gamma\right)  \theta_{0}}\dot{\theta}_{0}\text{, and
}r_{\mathrm{E}}\left(  \Gamma\right)  =\left(  \frac{2\Gamma}{\hslash}\right)
^{2}e^{-2\lambda\left(  \Gamma\right)  \theta_{0}}\dot{\theta}_{0}^{2}\text{,}
\label{speed4}%
\end{equation}
where $\lambda\left(  \Gamma\right)  \overset{\text{def}}{=}\left(
4\Gamma\right)  /h$. We observe that the IGC grows linearly in time with
$d\mathcal{C}_{\mathcal{M}_{s}}/d\tau\propto v_{\mathrm{E}}=r_{\mathrm{E}%
}^{1/2}$. \ In Fig. $1$, we compare the four driving schemes for relatively
small values of $\lambda$. We have three plots in Fig. $1$. In plot (a), we
represent the rescaled entropy production rate $\tilde{r}_{\mathrm{E}}$ with
$r_{\mathrm{E}}=\left(  2\Gamma/\hslash\right)  ^{2}\dot{\theta}_{0}^{2}%
\tilde{r}_{\mathrm{E}}$ as a function of $\lambda$. In plot (b), we depict the
behavior of the entropic efficiency $\eta_{\mathrm{E}}$ versus $\lambda$. In
plot (c), we represent the behavior of $\mathcal{\tilde{C}}_{\mathcal{M}_{s}}$
versus $\tau$. The quantity $\mathcal{\tilde{C}}_{\mathcal{M}_{s}}$ denotes
the rescaled version of the information geometric complexity $\mathcal{C}%
_{\mathcal{M}_{s}}$ in its long-time limit with $\mathcal{C}_{\mathcal{M}_{s}%
}^{\mathrm{asymptotic}}=\left(  \Gamma/\hslash\right)  \dot{\theta}%
_{0}\mathcal{\tilde{C}}_{\mathcal{M}_{s}}$. In (a), (b), and (c), the dotted,
dashed, thin solid, and thick solid lines correspond to the constant,
oscillatory, exponential law decay, and power law decay field intensity
behaviors, respectively. Finally, we set $\theta_{0}=1$ in all plots and
$\lambda=1/2$ in plot (c). Then, comparing Eqs. (\ref{speed3}) and
(\ref{speed4}), we arrive at the conclusion that for values of $\lambda$
sufficiently large this fourth scenario yields the coolest optimum paths that
are explored at the slowest entropic speed. In particular, when $\theta_{0}\in%
\mathbb{R}
_{+}$ and $\lambda\left(  \Gamma\right)  \overset{\text{def}}{=}\left(
4\Gamma\right)  /h\gg1$, the following chain of inequalities hold true,
\begin{equation}
0\leq e^{-\lambda\theta_{0}}\leq1/\left(  1+\lambda\theta_{0}\right)  ^{2}%
\leq\left\vert \cos\left(  \lambda\theta_{0}\right)  \right\vert \leq1\text{.}%
\end{equation}
Therefore, for values of the parameter $\lambda$ sufficiently large, the power
law decay strategy outruns the exponential decay strategy in terms of entropic
speed. To estimate numerically a typical value of $\lambda$ from a physics
standpoint, we recall that $\lambda=\left(  4\Gamma\right)  /h$,
$\Gamma=\left(  \left\vert e\right\vert \hslash B_{\bot}\right)  /2mc$ and,
thus, $\lambda=(1/\pi)\left(  \left\vert e\right\vert /mc\right)  $ $B_{\bot}%
$. Therefore, for a magnetic field with intensity $B_{\bot}$ of the order of
$0.1$ $\mathrm{T}$ (a half of a value typical of neodymium magnets),
$\lambda\approx18$ $\left[  \mathrm{MKSA}\right]  $. Interestingly, there are
parametric regions specified by smaller values of $\lambda$ (for instance,
$0\leq\lambda\lesssim1$) where the exponential-decay strategy can outperform
the power-law strategy in terms of entropic speed. However, its performance
declines in terms of either higher information geometric complexity or lower
entropic efficiency. In Fig. $2$, we have three plots. In plot (a), we depict
the entropy production rate $r_{\mathrm{E}}$ versus $\lambda$ and set
$\theta_{0}=1$. The thin solid and thick solid lines denote the exponential
law and the power law decay scenarios, respectively. The intersection between
the two lines occurs at $\lambda\simeq2.51$. In plot (b), we illustrate the
parametric region $\mathcal{D}\left(  \theta_{0}\text{, }\lambda\right)  $
where $r_{\mathrm{E}}^{\left(  \text{\textrm{exponential}}\right)  }\left(
\theta_{0}\text{, }\lambda\right)  \leq r_{\mathrm{E}}^{\left(
\text{\textrm{power-law}}\right)  }\left(  \theta_{0}\text{, }\lambda\right)
$ (black region). Finally, in plot (c) we visualize the ratios \textrm{R}%
$_{\mathcal{C}_{\mathcal{M}_{s}}}\overset{\text{def}}{=}\mathcal{C}%
_{\mathcal{M}_{s}}^{\left(  \text{\textrm{exponential}}\right)  }%
/\mathcal{C}_{\mathcal{M}_{s}}^{\left(  \text{\textrm{power-law}}\right)  }$
(thick solid line) and \textrm{R}$_{r_{\mathrm{E}}}\overset{\text{def}}%
{=}r_{\mathrm{E}}^{\left(  \text{\textrm{exponential}}\right)  }%
/r_{\mathrm{E}}^{\left(  \text{\textrm{power-law}}\right)  }$ (thin solid
line) versus $\lambda$ with $\theta_{0}$ set equal to one. We emphasize that
the exponential law decay scheme outperforms the power law decay scheme in
terms of both entropy production rate and information geometric complexity in
the limit of sufficiently large values of $\lambda$. As a final remark, we
remark that in all four scenarios it happens that $d\mathrm{\dot{C}%
}_{\mathcal{M}_{s}}/dv_{\mathrm{E}}=1/2\geq0$ with $\mathrm{\dot{C}%
}_{\mathcal{M}_{s}}\overset{\text{def}}{=}d\mathcal{C}_{\mathcal{M}_{s}}%
/d\tau$. Moreover, setting the efficiency $\eta_{\mathrm{E}}$ in Eq.
(\ref{efficiency}) equal to $\eta_{\mathrm{E}}\left(  r_{\min}\text{,
}r_{\mathrm{E}}\right)  $ with $r_{\mathrm{E}}=v_{\mathrm{E}}^{2}$, we have%
\begin{equation}
\frac{d\eta_{\mathrm{E}}}{dv_{\mathrm{E}}}=-\frac{4r_{\min}v_{\mathrm{E}}%
}{\left(  r_{\min}+v_{\mathrm{E}}^{2}\right)  ^{2}}\leq0\text{.}%
\end{equation}
Therefore, the entropic efficiency $\eta_{\mathrm{E}}$ is a monotonic
decreasing function of the entropic speed $v_{\mathrm{E}}$ while the temporal
rate of change of the information geometric complexity $\mathrm{\dot{C}%
}_{\mathcal{M}_{s}}$ is a monotonic increasing function of $v_{\mathrm{E}}$
with $d\eta_{\mathrm{E}}/dv_{\mathrm{E}}\leq0$ and $d\mathrm{\dot{C}%
}_{\mathcal{M}_{s}}/dv_{\mathrm{E}}\geq0$, respectively. A summary of the
relative ranking among the driving schemes considered appears in Table I.
\begin{table}[t]
\centering
\begin{tabular}
[c]{c|c|c|c}\hline\hline
\textbf{Hamiltonian Model} & \textbf{Rate of} \textbf{Entropy Production} &
\textbf{Efficiency} & \textbf{Complexity}\\\hline
$B_{\bot}$, constant & higher & lower & higher\\
$B_{\bot}$, oscillating & high & low & high\\
$B_{\bot}$, power law decay & low & high & low\\
$B_{\bot}$, exponential law decay & lower & higher & lower\\\hline
\end{tabular}
\caption{Schematic description of the entropy production rate $r_{\mathrm{E}}%
$, the entropic efficiency $\eta_{\mathrm{E}}$, and the information geometric
complexity $\mathcal{C}_{\mathcal{M}_{s}}$ in the four \textrm{su}$\left(
2;\mathbb{C}\right)  $ Hamiltonian being considered. In the limit of
sufficiently large values of the parameter $\lambda$ used to modify the
behavior of the external driving field, both higher efficiency and lower
complexity levels appear to be reached in the case of the driving scheme
specified by an exponential law decay.}%
\end{table}

\section{Concluding remarks}

We present here a summary of our main findings along with possible future directions.

\subsection{Summary of results}

We provided an information geometric description of quantum driving schemes
specified by $\mathrm{su}\left(  2\text{; }%
\mathbb{C}
\right)  $ time-dependent Hamiltonians (see Eq.\ (\ref{amo}) in terms of both
complexity (see Eq. (\ref{IGC})) and efficiency (see Eq. (\ref{efficiency}))
concepts. Specifically, starting from the parametrized pure output quantum
states $\left\{  \left\vert \psi\left(  \theta\right)  \right\rangle \right\}
$ describing the evolution of a spin-$1/2$ particle in an external
time-dependent magnetic field, we considered the probability paths $\left\{
p\left(  \theta\right)  \right\}  $ emerging from the parametrized squared
probability amplitudes of quantum origin with $\theta$ denoting statistical
parameter corresponding to the elapsed time. The information manifold
$\mathcal{M}_{s}$ of such paths was equipped with a Riemannian metrization
specified by the Fisher information $g_{\alpha\beta}\left(  \theta\right)  $
evaluated along the parametrized squared probability amplitudes. Employing a
minimum action principle, the optimum path connecting initial and final states
on the manifold in finite-time tuned out to be the geodesic path between the
two states. In particular, the total entropy production that occurs during the
transfer is minimized along these optimum paths (see Eqs. (\ref{geo1a}),
(\ref{geo2}), (\ref{geo3}), and (\ref{geo4})). For each optimum path that
emerges from the given quantum driving scheme, we evaluated (see Eqs.
(\ref{speed1}), (\ref{speed1b}), (\ref{speed3}), and (\ref{speed4})) the IGCs,
the entropic speeds, and the rates of entropy production used to define our
entropic efficiency measure in Eq. (\ref{efficiency}). From our analytical
estimates of complexity and efficiency, we provided a relative ranking among
the driving schemes being investigated (see Figure $1$, Figure $2$, and Table I).

The following points are of particular interest:

\begin{enumerate}
\item[{[i]}] We established a link between the IGC and the thermodynamic
length. Specifically, the IGC can be regarded as a measure of the
\textquotedblleft average\textquotedblright\ maximal number of statistically
distinguishable states along the path $\gamma_{\theta}$ since we have
$\mathcal{C}_{\mathcal{M}_{s}}\left(  \tau\right)  =\left\langle
\mathcal{L}\left(  \xi\right)  \right\rangle _{\tau_{0}\leq\xi\leq\tau}$. The
validity of this equation holds for the models we have investigated here.\ We
do not expect this relation to hold in its neat form in higher-dimensional
parameter spaces where, for instance, the equality between $ds\overset
{\text{def}}{=}\left[  g_{\alpha\beta}\left(  \theta\right)  d\theta^{\alpha
}d\theta^{\beta}\right]  ^{1/2}$ and $d\mathcal{V}\overset{\text{def}}%
{=}\left[  g\left(  \theta\right)  \right]  ^{1/2}d^{n}\theta$ with
$\theta=\left(  \theta^{1}\text{,..., }\theta^{n}\right)  $ does not hold
anylonger. After all, $\mathcal{C}_{\mathcal{M}_{s}}$ is related to volume
elements $d\mathcal{V}$ while $\mathcal{L}$ is a length emerging from the
integration of infinitesimal line elements $ds$.

\item[{[ii]}] We brought to light that the IGC of a geodesic path is connected
to the entropy production rate along that path. In particular, the rate of
change in time of the IGC is proportional to the square-root of the entropy
production rate along the path $\gamma_{\theta}$ where $\theta=\theta\left(
\xi\right)  $ with $\tau_{0}\leq\xi\leq\tau$, $d\mathcal{C}_{\mathcal{M}_{s}%
}/d\tau\propto r_{\mathrm{E}}^{1/2}$, with the constant of proportionality
being equal to $1/2$. The validity of this relation holds for the models we
considered. It would be interesting to explore what happens in more
complicated scenarios with a richer Hamiltonian dynamics with more tunable parameters.

\item[{[iii]}] We determined that the entropic efficiency $\eta_{\mathrm{E}}$
is a monotonic decreasing function of the entropic speed $v_{\mathrm{E}}$
while the temporal rate of change of the information geometric complexity
$\mathrm{\dot{C}}_{\mathcal{M}_{s}}\overset{\text{def}}{=}d\mathcal{C}%
_{\mathcal{M}_{s}}/d\tau$ is a monotonic increasing function of $v_{\mathrm{E}%
}$ with $d\eta_{\mathrm{E}}/dv_{\mathrm{E}}\leq0$ and $d\mathrm{\dot{C}%
}_{\mathcal{M}_{s}}/dv_{\mathrm{E}}\geq0$. Therefore, for the driving schemes
being considered here, higher speed values yield less efficient and more
complex probability paths. This is a manifestation of the so-called
speed-efficiency tradeoff along with the conjecture, at this stage, that
efficiency demands simplicity:\ less (complex) is more (efficient). A major
achievement would be that of constructing a driving scheme that is
simultaneously fast, efficient, and as simple as possible according to the
laws of physics. We believe the work presented here will help us pursue this
goal in future efforts.\textbf{ }
\end{enumerate}

\subsection{Outlook}

Our work can be improved in a number of ways. One of the main restrictions of
our investigation is its limitation to a single control parameter. However, we
believe our analysis can be extended to more than one control variable in a
relatively straightforward manner. Furthermore, our information geometric
analysis focuses on pure states and unitary evolution. In particular, we have
ignored considering more realistic scenarios where the quantum system is open
to the environment and dissipation effects in the form of dynamical
fluctuations of the controlled system (which, in general, is described by a
mixed quantum state) become important. In general, the temporal rate of change
of the density operator of an open quantum system can be expressed in terms of
the sum of two terms, the Hamiltonian piece and the dissipative piece
\cite{hayden21}. In a sense, the information geometric techniques we used here
can be regarded as applied to a closed quantum system viewed as an open
quantum system in the limit in which the dissipative piece is zero and the
rate of change of the density operator is solely expressed in terms of the
Hamiltonian piece. For a recent non-geometric study on the dynamics of a
two-level system which interacts with a dissipative bosonic environment at
zero temperature specified by a Lorentzian spectral density function, we refer
to Ref. \cite{wu21}. Moving from unitary evolution of pure states to open
systems described by impure states, a number of challenges emerge. For
instance, for pure states undergoing unitary dynamics, the Fisher information
metric is essentially the unique contractive Riemannian metric that can be
defined to quantify the distance between states \cite{brody19,guarnieri21}.
However, there is no unique suitable metric for characterizing the distance
between mixed states describing quantum systems open to the environment.
Furthermore, quantifying in an analytical manner minimum dissipation protocols
in the presence of a large number of experimentally tunable parameters is
rather challenging from a computational standpoint, in both classical and
quantum scenarios. For a numerically intensive investigation of nontrivial
minimum dissipation protocols for nanomagnetic (classical) spin models in the
presence of a large number of control parameters, we refer to Ref.
\cite{crooks17}. The extension of our proposed information geometric analysis
to the case of a dissipative dynamics of an open quantum system interacting
with an external environment in the presence of a large number of tunable
parameters can represent a number of additional challenges. They will be the
subject of forthcoming investigations.

In conclusion, despite its limitations, we believe that the analysis presented
here is a relevant piece of work that joins the increasing list of recent
investigations concerning an information geometric characterization of entropy
production and efficiency in both classical and quantum systems
\cite{hasegawa21,miller20,saito20,ito20} and deserves further investigation.

\begin{acknowledgments}
C.C. is grateful to the United States Air Force Research Laboratory (AFRL)
Summer Faculty Fellowship Program for providing support for this work. S.R.
acknowledges support from the National Research Council Research Associate
Fellowship program (NRC-RAP). P.M.A. acknowledges support from the Air Force
Office of Scientific Research (AFOSR). Any opinions, findings and conclusions
or recommendations expressed in this material are those of the author(s) and
do not necessarily reflect the views of the Air Force Research Laboratory
(AFRL). The Authors thank the anonymous Referees for stimulating comments
leading to an improved version of the manuscript.
\end{acknowledgments}

\bigskip\pagebreak

\appendix

\section{Equivalence of geodesic equations}

In this Appendix, motivated by Eq. (\ref{general1}) in\ Section III, we show
that the geodesic equations emerging from considering the variations
$\delta\left(  \int\sqrt{ds^{2}}\right)  $ and $\delta\left(  \int
ds^{2}\right)  $ with $ds^{2}\overset{\text{def}}{=}g_{\alpha\beta}\left(
\theta\right)  d\theta^{\alpha}d\theta^{\beta}$,%
\begin{equation}
\frac{d^{2}\theta^{\rho}}{d\xi^{2}}+\Gamma_{\mu\nu}^{\rho}\frac{d\theta^{\mu}%
}{d\xi}\frac{d\theta^{\nu}}{d\xi}=0\text{,} \label{geo1}%
\end{equation}
and,
\begin{equation}
\frac{d}{d\xi}\left(  g_{\mu\rho}\frac{d\theta^{\mu}}{d\xi}\right)  -\frac
{1}{2}\frac{d\theta^{\mu}}{d\xi}\frac{\partial g_{\mu\nu}}{\partial\xi^{\rho}%
}\frac{d\theta^{\nu}}{d\xi}=0\text{,}%
\end{equation}
respectively, are equivalent. Indeed, using standard tensor algebra
techniques, observe that%
\begin{align}
0  &  =\frac{d}{d\xi}\left(  g_{\mu\rho}\frac{d\theta^{\mu}}{d\xi}\right)
-\frac{1}{2}\frac{d\theta^{\mu}}{d\xi}\frac{\partial g_{\mu\nu}}%
{\partial\theta^{\rho}}\frac{d\theta^{\nu}}{d\xi}\nonumber\\
&  =\frac{d}{d\xi}\left(  g_{\mu\rho}\right)  \frac{d\theta^{\mu}}{d\xi
}+g_{\mu\rho}\frac{d^{2}\theta^{\mu}}{d\xi^{2}}-\frac{1}{2}\frac{\partial
g_{\mu\nu}}{\partial\theta^{\rho}}\frac{d\theta^{\mu}}{d\xi}\frac{d\theta
^{\nu}}{d\xi}\nonumber\\
&  =\frac{\partial g_{\mu\rho}}{\partial\theta^{\nu}}\frac{d\theta^{\nu}}%
{d\xi}\frac{d\theta^{\mu}}{d\xi}+g_{\mu\rho}\frac{d^{2}\theta^{\mu}}{d\xi^{2}%
}-\frac{1}{2}\frac{\partial g_{\mu\nu}}{\partial\theta^{\rho}}\frac
{d\theta^{\mu}}{d\xi}\frac{d\theta^{\nu}}{d\xi}\nonumber\\
&  =\frac{1}{2}\left(  \frac{\partial g_{\mu\rho}}{\partial\theta^{\nu}}%
+\frac{\partial g_{\nu\rho}}{\partial\theta^{\mu}}\right)  \frac{d\theta^{\mu
}}{d\xi}\frac{d\theta^{\nu}}{d\xi}+g_{\mu\rho}\frac{d^{2}\theta^{\mu}}%
{d\xi^{2}}-\frac{1}{2}\frac{\partial g_{\mu\nu}}{\partial\theta^{\rho}}%
\frac{d\theta^{\mu}}{d\xi}\frac{d\theta^{\nu}}{d\xi}\nonumber\\
&  =g_{\mu\rho}\frac{d^{2}\theta^{\mu}}{d\xi^{2}}+\frac{1}{2}\left(
\frac{\partial g_{\mu\rho}}{\partial\theta^{\nu}}+\frac{\partial g_{\nu\rho}%
}{\partial\theta^{\mu}}-\frac{\partial g_{\mu\nu}}{\partial\theta^{\rho}%
}\right)  \frac{d\theta^{\mu}}{d\xi}\frac{d\theta^{\nu}}{d\xi}\nonumber\\
&  =g_{\mu\rho}\frac{d^{2}\theta^{\mu}}{d\xi^{2}}+\Gamma_{\rho\text{, }\mu\nu
}\frac{d\theta^{\mu}}{d\xi}\frac{d\theta^{\nu}}{d\xi}\nonumber\\
&  =g^{\rho\rho}g_{\mu\rho}\frac{d^{2}\theta^{\mu}}{d\xi^{2}}+g^{\rho\rho
}\Gamma_{\rho\text{, }\mu\nu}\frac{d\theta^{\mu}}{d\xi}\frac{d\theta^{\nu}%
}{d\xi}\nonumber\\
&  =\frac{d^{2}\theta^{\rho}}{d\xi^{2}}+\Gamma_{\mu\nu}^{\rho}\frac
{d\theta^{\mu}}{d\xi}\frac{d\theta^{\nu}}{d\xi}\text{.}%
\end{align}

Therefore, we conclude that Eqs. (\ref{geo1}) and (\ref{geo2}) are equivalent.

\section{Physical significance of entropy production rate}

In this Appendix, we discuss the physical significance of the concept of
entropy production rate in Eq. (\ref{EPR}) of Section III in relation to the
thermodynamics of a system of spin-$1/2$ particles driven by an external
magnetic field.

\subsection{Negative temperature of a spin-$1/2$ particle in an external
magnetic field}

In statistical physics, phenomena of negative temperatures have much less
practical importance than phenomena of positive temperatures. However,
negative temperatures are characteristic of atomic systems with inverted
populations, and they can be equally well-described from a thermodynamical
standpoint. In particular, if the entropy $\sigma$ of a system is not a
monotonically increasing function of its internal energy $U$, it exhibits a
negative temperature whenever $1/T\overset{\text{def}}{=}\left(
\partial\sigma/\partial U\right)  _{X}$ $\ $is negative with $X$ standing for
all the other extensive variables the entropy might depend upon
\cite{kittel58}. More generally, the three essential requirements for a
thermodynamical system to be capable of negative temperature are
\cite{ramsey56}: i) In order to describe the system in terms of the concept of
temperature, the elements of the system must be in thermodynamical equilibrium
among themselves; ii) There must be an upper bound to the values of the energy
of the allowed states of the system; iii) The system must be thermally
isolated from all systems which do not fulfill conditions i) and ii). A simple
physical example of a system for which a negative temperature emerges is given
by a spin-$1/2$ particles in an external magnetic field with only two energy
states available to each element of the system. Let us assume that the
energies of the upper and lower states are given by $\epsilon_{2}%
\overset{\text{def}}{=}+\epsilon$ and $\epsilon_{1}\overset{\text{def}}%
{=}-\epsilon$, respectively, so that the energy gap between the two energy
levels is $\Delta\epsilon\overset{\text{def}}{=}\epsilon_{2}-\epsilon
_{1}=2\epsilon>0$. Furthermore, let us denote with $p_{\epsilon_{2}}$ and
$p_{\epsilon_{1}}$ the probabilities of occupying the upper and lower states,
respectively, with $p_{\epsilon_{2}}+p_{\epsilon_{1}}=1$. Making use of the
canonical ensemble formalism in statistical mechanics, setting the Boltzmann
constant $k_{B}$ equal to one, and recalling that the entropy of the system is
the logarithm of the number of accessible states, it happens that
$\sigma\left(  U\right)  $ can be recast as%
\begin{equation}
\sigma\left(  U\right)  =-\left[  \frac{N\epsilon_{2}-U}{N\Delta\epsilon}%
\log\left(  \frac{N\epsilon_{2}-U}{N\Delta\epsilon}\right)  +\frac
{U-N\epsilon_{1}}{N\Delta\epsilon}\log\left(  \frac{U-N\epsilon_{1}}%
{N\Delta\epsilon}\right)  \right]  \text{,} \label{entropySM}%
\end{equation}
where $-N\epsilon\leq U\overset{\text{def}}{=}\sum_{i}p_{\epsilon_{i}}%
\epsilon_{i}=\langle E\rangle\leq N\epsilon$ is the average energy of the
system, with $N$ denoting the total number of elements of the system. The
region of negative slope of this curve $\sigma\left(  U\right)  $ in Eq.
(\ref{entropySM}) corresponds to negative temperature. When the lowest
possible energy state is fully occupied, we have $p_{\epsilon_{1}}=1$,
$U=-N\epsilon$, and the state is a highly ordered state at $+0^{\circ
}\mathrm{K}$ with $\sigma=0$. When the highest possible energy state is fully
occupied, instead, we have $p_{\epsilon_{2}}=1$, $U=+N\epsilon$, and the state
is a highly ordered state at $-0^{\circ}\mathrm{K}$ with $\sigma=0$. The
states at $\pm0^{\circ}\mathrm{K}$ are completely different from a physics
standpoint. When the system is at $+0^{\circ}\mathrm{K}$, it cannot become
colder since it cannot give up its energy anymore. When the system is at
$-0^{\circ}\mathrm{K}$, instead, it cannot become hotter since it cannot
absorb energy anymore. We remark that a system in a negative temperature state
is very hot and gives up energy to any system at positive temperature put into
contact with it. Negative temperatures correspond to higher energies than
positive temperatures. Furthermore, unlike what happens for positive
temperatures, an increased internal energy corresponds to diminished entropy
at negative temperatures. At intermediate energies with $-N\epsilon
<U<N\epsilon$, when some elements are in the low-energy state and others in
the high-energy state, there is greater entropy since there is less order.
Therefore, between the lowest and the highest energy states of the
thermodynamic system, the entropy passes through a maximum and then diminishes
with increasing $U$. The maximum (with $\sigma_{\max}=\log2$) occurs at $U=0$
where $p_{\epsilon_{1}}=p_{\epsilon_{2}}=1/2$. Right before and after $U=0$,
$T=+\infty^{\circ}\mathrm{K}$ and $T=-\infty^{\circ}\mathrm{K}$, respectively.
This change of sign in the temperature is a consequence of the inversion of
the population levels with $p_{\epsilon_{2}}/p_{\epsilon_{1}}=e^{-\beta
\Delta\epsilon}>1$ when $0<U<+N\epsilon$ with $\beta\overset{\text{def}}%
{=}\left(  k_{B}T\right)  ^{-1}$. For a discussion on experimental
realizations of negative temperatures with systems of interacting nuclear
spins, we refer to Refs. \cite{purcell51,proctor57}.

\subsection{Physical interpretation of the rate of entropy production}

In what follows, we provide a more physical interpretation of the rate of
entropy production given by,%
\begin{equation}
r_{\mathrm{E}}\overset{\text{def}}{=}\frac{d}{d\tau}\mathcal{I}(\tau)=\frac
{d}{d\tau}\left[  \int_{0}^{\tau}\frac{d\theta^{\alpha}}{d\xi}g_{\alpha\beta
}\left(  \theta\right)  \frac{d\theta^{\beta}}{d\xi}d\xi\right]  \text{,}
\label{ER1}%
\end{equation}
where $\mathcal{I}(\tau)$ is the \textit{thermodynamic divergence}, by
exploiting our thermodynamic considerations concerning negative temperature
spin-$1/2$ systems in external magnetic fields with entropy given as in Eq.
(\ref{entropySM}).

For a physical system in equilibrium with a large thermal reservoir, it
happens that the thermodynamic metric tensor $g_{\alpha\beta}\left(
\theta\left(  \xi\right)  \right)  $ in Eq. (\ref{ER1}) represents the
covariance matrix of fluctuations around equilibrium,
\begin{equation}
\overline{\delta X}_{\alpha\beta}^{2}\overset{\text{def}}{=}\left\langle
\left(  X_{\alpha}-\left\langle X_{\alpha}\right\rangle \right)  \left(
X_{\beta}-\left\langle X_{\beta}\right\rangle \right)  \right\rangle \text{,}%
\end{equation}
with $\left\{  X_{\alpha}\left(  x\right)  \right\}  $ being the thermodynamic
variables that specify the Hamiltonian of the system while $\left\{
x\right\}  $ are the configuration space variables. Moreover, $\left\{
\theta^{\alpha}\right\}  $ are the experimentally controllable parameters of
the system and $\left\langle \cdot\right\rangle $ denotes the ensemble average
with respect to the canonical Gibbs distribution function $p\left(
x|\theta\right)  \equiv p_{x}\left(  \theta\right)  =e^{-\theta^{\alpha
}\left(  \xi\right)  X_{\alpha}\left(  x\right)  }/\mathcal{Z}$ with
$\mathcal{Z}$ being the partition function of the system. After some
straightforward algebra, it can be shown that the thermodynamic metric tensor
and the Fisher-Rao information metric tensor are equivalent. Specifically, we
have%
\begin{equation}
g_{\alpha\beta}\left(  \theta\right)  =\overline{\delta X}_{\alpha\beta}^{2}=%
{\displaystyle\sum\limits_{x}}
p_{x}\left(  \theta\right)  \frac{\partial\log p_{x}(\theta)}{\partial
\theta^{\alpha}}\frac{\partial\log p_{x}(\theta)}{\partial\theta^{\beta}%
}\text{.} \label{ER2}%
\end{equation}
Using Eq. (\ref{ER2}), $r_{\mathrm{E}}$ in Eq. (\ref{ER1}) can be recast as%
\begin{equation}
r_{\mathrm{E}}=\frac{d\theta^{\alpha}}{d\xi}\overline{\delta X}_{\alpha\beta
}^{2}\frac{d\theta^{\beta}}{d\xi}=%
{\displaystyle\sum\limits_{x}}
p_{x}(\theta)\left(  \frac{d\log p_{x}(\theta)}{d\xi}\right)  ^{2}\text{,}
\label{ER3}%
\end{equation}
with $\theta=\theta\left(  \xi\right)  \overset{\text{def}}{=}\left(
\theta^{1}\left(  \xi\right)  \text{,..., }\theta^{n}\left(  \xi\right)
\right)  $ with $n$ being the dimensionality of the parameter space. From Eqs.
(\ref{ER2}) and (\ref{ER3}), we point out that while the Fisher-Rao
information metric tensor is defined in terms of \textit{partial} derivatives
of the probabilities with respect to the control parameters, the rate of
entropy production is specified by \textit{total} derivative of the
probabilities $\tfrac{dp_{x}(\theta)}{d\xi}=\tfrac{\partial p_{x}(\theta
)}{\partial\theta^{\alpha}}\tfrac{d\theta^{\alpha}}{d\xi}$, with respect to
the affine parameter $\xi$ along the trajectories $\left\{  \theta^{\alpha
}\left(  \xi\right)  \right\}  $. For a \textit{single} control parameter
$\theta^{\alpha}(\xi)\rightarrow\theta(\xi)$ the above two formulas Eq.
(\ref{ER2}) and Eq. (\ref{ER3}) are deceptively similar
\begin{equation}
g_{\alpha\beta}(\theta)\rightarrow g(\theta)=\sum_{x}p_{x}(\theta)\left(
\frac{\partial\log p_{x}(\theta)}{\partial\theta}\right)  ^{2}\text{,}\quad
r_{\mathrm{E}}\rightarrow\sum_{x}p_{x}(\theta)\left(  \frac{d\log p_{x}%
(\theta)}{d\xi}\right)  ^{2}\text{,} \label{newpaul}%
\end{equation}
differing only in the use of type of derivative employed to differentiate the
\textquotedblleft score function\textquotedblright\ $\log p_{x}(\theta)$.
Moreover, from Eq. (\ref{ER3}), we observe that $r_{\mathrm{E}}$ can be also
described as the \textquotedblleft product\textquotedblright\ of the
fluctuation term $\overline{\delta X}_{\alpha\beta}^{2}$ and the square of the
total rate of change with respect to the affine parameter $\xi$ of the control
parameter $\theta^{\alpha}\left(  \xi\right)  $.

To better grasp the significance of the rate of entropy production, we
consider the following illustrative comparison. First, we consider a system
with probability path defined by $p\left(  \theta\right)  \overset{\text{def}%
}{=}\left(  p_{w}\left(  \theta\right)  \text{, }p_{w_{\perp}}\left(
\theta\right)  \right)  =\left(  \sin^{2}\left(  \theta\right)  \text{, }%
\cos^{2}\left(  \theta\right)  \right)  $ with $\theta\left(  \xi\right)
=\left(  \pi/2\right)  \xi$ and $0\leq\xi\leq1$. In this first case, we obtain
$r_{\mathrm{E}}=\pi^{2}$. This system exhibits features that are similar to
those characterizing a two-level system as transparent from the link between
the probabilities defining $p\left(  \theta\right)  \overset{\text{def}}%
{=}\left(  p_{w}\left(  \theta\right)  \text{, }p_{w_{\perp}}\left(
\theta\right)  \right)  \rightarrow$ $p\left(  \beta\right)  \overset
{\text{def}}{=}\left(  p_{\epsilon_{1}}\left(  \beta\right)  \text{,
}p_{\epsilon_{2}}\left(  \beta\right)  \right)  $. Here, as in Crooks
\cite{crooks07,crooks12}, we use the variable\textbf{ }inverse temperature
$\beta(\xi)=\left[  k_{B}T(\xi)\right]  ^{-1}$ as the single control parameter
$\theta$. For the canonical ensemble we have $p_{\epsilon_{i}}\left(
\beta\right)  \overset{\text{def}}{=}e^{-\beta\epsilon_{i}}/\mathcal{Z}%
=e^{-(\beta\epsilon_{i}+\log\mathcal{Z})}$ with $\mathcal{Z}\overset
{\text{def}}{=}e^{-\beta\epsilon_{1}}+e^{-\beta\epsilon_{2}}$. Noting that
$\frac{\partial\log p_{\epsilon_{i}}}{d\beta}=\overline{\epsilon}-\epsilon
_{i}$, with $\overline{\epsilon}\overset{\text{def}}{=}\sum_{i}p_{\epsilon
_{i}}(\beta)\epsilon_{i}$ we see that the Fisher information in
Eq.(\ref{newpaul}) becomes $g(\theta)=\overline{\delta E}^{2}\overset
{\text{def}}{=}\sum_{i}p_{\epsilon_{i}}\left(  \beta\right)  \left(
\epsilon_{i}-\bar{\epsilon}\right)  ^{2}$, the variance of the energy
fluctuations. Inserting this into the expression for $r_{\text{E}}$ in
Eq.(\ref{ER3}), and using $\tfrac{dp_{x}(\theta)}{d\xi}=\tfrac{\partial
p_{x}(\theta)}{\partial\beta}\tfrac{d\beta}{d\xi}$ we find that the entropy
production rate $r_{\mathrm{E}}$ is the product of the energy fluctuations
$\overline{\delta E}^{2}$ times the squared rate of change $\left(
\tfrac{d\beta}{d\xi}\right)  ^{2}$ of the control parameter along the
trajectory in parameter space
\begin{equation}
r_{\mathrm{E}}=\overline{\delta E}^{2}\left(  \frac{d\beta}{d\xi}\right)
^{2}=\left(  \frac{dp_{\epsilon_{1}}}{d\xi}\right)  ^{2}\left(  \frac
{1}{p_{\epsilon_{1}}}+\frac{1}{p_{\epsilon_{2}}}\right)  =\pi^{2}\text{.}
\label{ER4}%
\end{equation}
Note that the second equality in Eq.(\ref{ER4}) arises (after a little
algebra) from comparing the two terms in the first equality, respectively, to
functions of the probabilities. Eq. (\ref{ER4}) is in agreement with the
calculation of the rate of entropy production carried out in the first case.
Moreover, we remark that the quantity $dp_{\epsilon_{1}}/d\xi$, which acts as
a kind of \textquotedblleft probability velocity\textquotedblright\ for the
two level system along the trajectory in parameter space, can be regarded as a
relative energy fluctuation term since $dp_{\epsilon_{1}}/d\xi\propto
\overline{\delta E}/\epsilon$ while the term $\left(  1/p_{\epsilon_{1}%
}+1/p_{\epsilon_{2}}\right)  $ (which, we note, can be viewed as the
reciprocal of a reduced probability mass term) is proportional to the square
of the rate of change of the control parameter $\beta$ with respect to the
affine time parameter $\xi$, $\left(  1/p_{\epsilon_{1}}+1/p_{\epsilon_{2}%
}\right)  \propto\left(  d\beta/d\xi\right)  ^{2}$. Thus, the form of
Eq.(\ref{ER4}) is reminiscent of a kind of \textquotedblleft kinetic energy of
fluctuations\textquotedblright\ of the two level system along the trajectory
in parameter space. As a side remark, we refer to Ref. \cite{adolfo18} for an
interesting link between the Fisher information function in information
geometry and the concept of entropic acceleration in thermodynamics. Finally,
for further details on negative temperatures and fluctuations in
thermodynamics, we refer to Ref. \cite{kittel58}.

\section{Preserving the relative ranking}

In this Appendix, we check that the relative ranking of the driving schemes
provided by the three distinct measures of entropic efficiency introduced in
Section III is preserved.

Recall that the three efficiency measures $\eta_{\mathrm{E}}^{\left(
1\right)  }\left(  r_{\mathrm{E}}\right)  $, $\eta_{\mathrm{E}}^{\left(
2\right)  }\left(  r_{\mathrm{E}}\right)  $, and $\eta_{\mathrm{E}}\left(
r_{\mathrm{E}}^{\left(  l\right)  }\text{, }r_{\mathrm{E}}^{\left(  m\right)
}\right)  $ are defined as,%
\begin{equation}
\eta_{\mathrm{E}}^{\left(  1\right)  }\left(  r_{\mathrm{E}}\right)
\overset{\text{def}}{=}1-\frac{r_{\mathrm{E}}}{r_{\mathrm{E}}^{\max}}\text{,
}\eta_{\mathrm{E}}^{\left(  2\right)  }\left(  r_{\mathrm{E}}\right)
\overset{\text{def}}{=}\frac{r_{\mathrm{E}}^{\min}}{r_{\mathrm{E}}}\text{, and
}\eta_{\mathrm{E}}\left(  r_{\mathrm{E}}^{\left(  l\right)  }\text{,
}r_{\mathrm{E}}^{\left(  m\right)  }\right)  \overset{\text{def}}{=}%
1-\frac{\left\vert r_{\mathrm{E}}^{\left(  l\right)  }-r_{\mathrm{E}}^{\left(
m\right)  }\right\vert }{r_{\mathrm{E}}^{\left(  l\right)  }+r_{\mathrm{E}%
}^{\left(  m\right)  }}\text{,}%
\end{equation}
respectively. If we assume that $r_{\mathrm{E}}^{\left(  i_{\ast}\right)
}\geq r_{\mathrm{E}}^{\left(  i_{\ast}^{\prime}\right)  }$, a straightforward
calculation yields
\begin{equation}
\eta_{\mathrm{E}}^{\left(  1\right)  }\left(  r_{\mathrm{E}}^{\left(  i_{\ast
}\right)  }\right)  \leq\eta_{\mathrm{E}}^{\left(  1\right)  }\left(
r_{\mathrm{E}}^{\left(  i_{\ast}^{\prime}\right)  }\right)  \text{, }%
\eta_{\mathrm{E}}^{\left(  2\right)  }\left(  r_{\mathrm{E}}^{\left(  i_{\ast
}\right)  }\right)  \leq\eta_{\mathrm{E}}^{\left(  2\right)  }\left(
r_{\mathrm{E}}^{\left(  i_{\ast}^{\prime}\right)  }\right)  \text{, and }%
\eta_{\mathrm{E}}\left(  r_{\mathrm{E}}^{\left(  i_{\ast}\right)  }\text{,
}r_{\mathrm{E}}^{\min}\right)  \leq\eta_{\mathrm{E}}\left(  r_{\mathrm{E}%
}^{\left(  i_{\ast}^{\prime}\right)  }\text{, }r_{\mathrm{E}}^{\min}\right)
\text{.} \label{yo1}%
\end{equation}
Similarly, when $r_{\mathrm{E}}^{\left(  i_{\ast}\right)  }\leq r_{\mathrm{E}%
}^{\left(  i_{\ast}^{\prime}\right)  }$, we get%
\begin{equation}
\eta_{\mathrm{E}}^{\left(  1\right)  }\left(  r_{\mathrm{E}}^{\left(  i_{\ast
}\right)  }\right)  \geq\eta_{\mathrm{E}}^{\left(  1\right)  }\left(
r_{\mathrm{E}}^{\left(  i_{\ast}^{\prime}\right)  }\right)  \text{, }%
\eta_{\mathrm{E}}^{\left(  2\right)  }\left(  r_{\mathrm{E}}^{\left(  i_{\ast
}\right)  }\right)  \geq\eta_{\mathrm{E}}^{\left(  2\right)  }\left(
r_{\mathrm{E}}^{\left(  i_{\ast}^{\prime}\right)  }\right)  \text{, and }%
\eta_{\mathrm{E}}\left(  r_{\mathrm{E}}^{\left(  i_{\ast}\right)  }\text{,
}r_{\mathrm{E}}^{\min}\right)  \geq\eta_{\mathrm{E}}\left(  r_{\mathrm{E}%
}^{\left(  i_{\ast}^{\prime}\right)  }\text{, }r_{\mathrm{E}}^{\min}\right)
\text{.} \label{yo2}%
\end{equation}
Note that $i_{\ast}$ and $i_{\ast}^{\prime}$ are arbitrary indices with $1\leq
i_{\ast}$, $i_{\ast}^{\prime}\leq\bar{N}$ with $\bar{N}$ being the number of
different driving schemes being ranked. Therefore, given the arbitrariness of
the inequalities $r_{\mathrm{E}}^{\left(  i_{\ast}\right)  }\geq
r_{\mathrm{E}}^{\left(  i_{\ast}^{\prime}\right)  }$ and $r_{\mathrm{E}%
}^{\left(  i_{\ast}\right)  }\leq r_{\mathrm{E}}^{\left(  i_{\ast}^{\prime
}\right)  }$, we conclude from Eqs. (\ref{yo1}) and (\ref{yo2}) that
$\eta_{\mathrm{E}}^{\left(  1\right)  }\left(  r_{\mathrm{E}}\right)  $,
$\eta_{\mathrm{E}}^{\left(  2\right)  }\left(  r_{\mathrm{E}}\right)  $, and
$\eta_{\mathrm{E}}\left(  r_{\mathrm{E}}^{\left(  l\right)  }\text{,
}r_{\mathrm{E}}^{\left(  m\right)  }\right)  $ rank the driving schemes in a
similar manner by preserving the relative order from the best one to the worst one.

\section{Parametrization of probability amplitudes}

In this Appendix, we report a general parametrization of the probability
amplitudes $\alpha\left(  t\right)  $ and $\beta\left(  t\right)  $ that
appear in Eq. (\ref{good}) of Section IV.

Following Refs. \cite{messina14,grimaudo18}, it happens that $\alpha\left(
t\right)  $ and $\beta\left(  t\right)  $ in Eq. (\ref{good}) can be recast
as,%
\begin{equation}
\left\{
\begin{array}
[c]{c}%
\alpha\left(  t\right)  =\left\{  \cos\left[  \Phi\left(  t\right)  \right]
-i\frac{b}{\sqrt{1+b^{2}}}\sin\left[  \Phi\left(  t\right)  \right]  \right\}
e^{i\frac{\phi_{\omega}\left(  t\right)  }{2}}\text{,}\\
\\
\beta\left(  t\right)  =\frac{1}{\sqrt{1+b^{2}}}\sin\left[  \Phi\left(
t\right)  \right]  e^{i\left[  \frac{\phi_{\omega}\left(  t\right)  }{2}%
-\frac{\pi}{2}\right]  }\text{,}%
\end{array}
\right.  \label{gigi}%
\end{equation}
with $\Phi\left(  t\right)  $ defined as $\Phi\left(  t\right)  \overset
{\text{def}}{=}\sqrt{1+b^{2}}\int_{0}^{t}\frac{\left\vert \omega\left(
t^{\prime}\right)  \right\vert }{\hslash}dt^{\prime}$, provided that
$\Omega\left(  t\right)  +\hslash\dot{\phi}_{\omega}\left(  t\right)
/2=b\left\vert \omega\left(  t\right)  \right\vert $ where $b$ is an arbitrary
real number/parameter. Clearly, $\left\vert \omega\left(  t\right)
\right\vert $ is the magnitude of the complex transverse field $\omega\left(
t\right)  =\left\vert \omega\left(  t\right)  \right\vert e^{i\phi_{\omega
}\left(  t\right)  }$ with $\left\vert \omega\left(  t\right)  \right\vert
\propto B_{\perp}\left(  t\right)  $ and $\Omega\left(  t\right)  $ is the
real longitudinal field with $\left\vert \Omega\left(  t\right)  \right\vert
\propto B_{\parallel}\left(  t\right)  $. In particular, the resonance regime
is specified by $b\rightarrow0$ with $\alpha\left(  t\right)  $ and
$\beta\left(  t\right)  $ in Eq. (\ref{gigi}) reducing to%
\begin{equation}
\left\{
\begin{array}
[c]{c}%
\alpha\left(  t\right)  =\cos\left[  \int_{0}^{t}\frac{\left\vert
\omega\left(  t^{\prime}\right)  \right\vert }{\hslash}dt^{\prime}\right]
e^{i\frac{\phi_{\omega}\left(  t\right)  }{2}}\text{,}\\
\beta\left(  t\right)  =\sin\left[  \int_{0}^{t}\frac{\left\vert \omega\left(
t^{\prime}\right)  \right\vert }{\hslash}dt^{\prime}\right]  e^{i\left[
\frac{\phi_{\omega}\left(  t\right)  }{2}-\frac{\pi}{2}\right]  }\text{.}%
\end{array}
\right.  \label{gigi2}%
\end{equation}
For more details, we refer to the original work in Refs.
\cite{messina14,grimaudo18}.


\begin{thebibliography}{99}                                                                                               %


\bibitem {kronsjo87}L. Kronsj\"{o}, \emph{Algorithms: Their Complexity and
Efficiency}, John Wiley \& Sons Ltd. (1987).

\bibitem {montanaro16}A. Montanaro, \emph{Quantum algorithms: An overview},
npj Quantum Information 2, 15023 (2016).

\bibitem {nielsen}M. A. Nielsen and I. L. Chuang, \emph{Quantum Computation
and Quantum Information}, Cambridge University Press (2000).

\bibitem {cafaro11}C. Cafaro and S. Mancini, \emph{A geometric algebra
perspective on quantum computational gates and universality in quantum
computing}, Advances in Applied Clifford Algebras \textbf{21}, 493 (2011).

\bibitem {kung73}H. T. Kung and J. F. Traub, \emph{Computational complexity of
one-point and multi-point iteration}, in \emph{Complexity of Real
Computation}, R. Karp, Editor, American Mathematical Society, Providence, RI (1973).

\bibitem {amari}S. Amari and \ H. Nagaoka, \emph{Methods of Information
Geometry}, Oxford University Press (2000).

\bibitem {ruppeiner95}G. Ruppeiner, \emph{Riemannian geometry in thermodynamic
fluctuation theory}, Rev. Mod. Phys.\textbf{\ 67}, 605 (1995).

\bibitem {ruppeiner96}G. Ruppeiner, \emph{Erratum: Riemannian geometry in
thermodynamic fluctuation theory}, Rev. Mod. Phys. \textbf{68}, 313 (1996).

\bibitem {hasegawa21}T. Van Vu and Y. Hasegawa, \emph{Geometrical bounds of
the irreversibility in Markovian systems}, Phys. Rev. Lett. \textbf{126},
010601 (2021).

\bibitem {miller20}H. J. D. Miller and M. Mehboudi, \emph{Geometry of work
fluctuations versus efficiency in microscopic thermal machines}, Phys. Rev.
Lett. \textbf{125}, 260602 (2020).

\bibitem {saito20}K. Brandner and K. Saito, \emph{Thermodynamic geometry of
microscopic heat engines}, Phys. Rev. Lett. \textbf{124}, 040602 (2020).

\bibitem {ito20}S. Ito, M. Oizumi, and S.\ Amari, \emph{Unified framework for
the entropy production and the stochastic interaction based on information
geometry}, Phys. Rev. Research \textbf{2}, 033048 (2020).

\bibitem {cafaro18a}C. Cafaro and P. M.\ Alsing, \emph{Decrease of Fisher
information and the information geometry of evolution equations for quantum
mechanical probability amplitudes}, Phys. Rev. \textbf{E97}, 042110 (2018).

\bibitem {cafaro20}C. Cafaro and P. M.\ Alsing, \emph{Information geometry
aspects of minimum entropy production paths from quantum mechanical
evolutions}, Physical Review \textbf{E101}, 022110 (2020).

\bibitem {gassner21}S. Gassner, C. Cafaro, S. A. Ali, and P. M. Alsing,
\emph{Information geometric aspects of probability paths with minimum entropy
production for quantum state evolution}, Int. J. Geom. Meth. Mod. Phys.
\textbf{18}, 2150127 (2021).

\bibitem {alsing19}C. Cafaro and P. M. Alsing, \emph{Continuous-time quantum
search and time-dependent two-level quantum systems}, Int. J. Quantum
Information \textbf{17}, 1950025 (2019).

\bibitem {messina14}A. Messina and H. Nakazato, \emph{Analytically solvable
Hamiltonians for quantum two-level systems and their dynamics}, J. Phys. A:
Math. Theor. \textbf{47}, 445302 (2014).

\bibitem {grimaudo18}R. Grimaudo, A. S. M. de Castro, H. Nakazato, and A.
Messina, \emph{Classes of exactly solvable generalized semi-classical Rabi
systems}, Ann. Phys. (Berlin) \textbf{2018}, 1800198.

\bibitem {castelvecchi17}D. Castelvecchi, \emph{Clash of the physics laws},
Nature (London) 543, 597 (2017).

\bibitem {braunstein96}S. L. Braunstein, C. M. Caves, and G. J. Milburn,
\emph{Generalized uncertainty relations: Theory, examples, and Lorentz
invariance}, Ann. Phys. \textbf{247}, 135 (1996).

\bibitem {cafaro07}C. Cafaro and S. A. Ali, \emph{Jacobi fields on statistical
manifolds of negative curvature}, Physica \textbf{D234}, 70 (2007).

\bibitem {cafarothesis}C. Cafaro, \emph{The information geometry of chaos},
PhD Thesis, State University of New York at Albany, Albany-NY, USA (2008).
Available online at arXiv: math-ph/1601.07935 (2016).

\bibitem {ali18}S. A. Ali, C. Cafaro, S. Gassner, and A. Giffin, \emph{An
information geometric perspective on the complexity of macroscopic predictions
arising from incomplete information}, Adv. Math. Phys., Volume 2018, Article
ID 2048521 (2018).

\bibitem {ali10}S. A. Ali, C. Cafaro, D.-H. Kim, and S. Mancini, \emph{The
effect of microscopic correlations on the information geometric complexity of
Gaussian statistical models}, Physica \textbf{A389}, 3117 (2010).

\bibitem {cafaro18}D. Felice, C. Cafaro, and S. Mancini, \emph{Information
geometric methods for complexity}, Chaos \textbf{28}, 032101 (2018).

\bibitem {cafaro17}S. A.\ Ali and C. Cafaro, \emph{Theoretical investigations
of an information geometric approach to complexity}, Rev. Math. Phys.
\textbf{29}, 1730002 (2017).\emph{\ }

\bibitem {cafaro10}C. Cafaro, A. Giffin, S. A. Ali, and D.-H. Kim,
\emph{Reexamination of an information geometric construction of entropic
indicators of complexity}, Appl. Math. Comput. \textbf{217}, 2944 (2010).

\bibitem {crooks07}G. E. Crooks,\emph{\ Measuring thermodynamic length}, Phys.
Rev. Lett. \textbf{99}, 100602 (2007).

\bibitem {ruppeiner79}G. Ruppeiner, \emph{Thermodynamics: A Riemannian
geometric model}, Phys. Rev. \textbf{A20}, 1608 (1979).

\bibitem {weinhold75}F. Weinhold, \emph{Metric geometry of equilibrium
thermodynamics}, J. Chem. Phys. \textbf{63}, 2479 (1975).

\bibitem {salamon83}P. Salamon and R. S. Berry, \emph{Thermodynamic length and
dissipated availability}, Phys. Rev.\ Lett. \textbf{51}, 1127 (1983).

\bibitem {feldmann85}T. Feldmann, B. Andresen, A.\ Qi, and P. Salamon,
\emph{Thermodynamic lengths and intrinsic time scales in molecular
relaxation}, J. Chem. Phys. \textbf{83}, 5849 (1985).

\bibitem {wootters81}W. K. Wootters, \emph{Statistical distance and Hilbert
space}, Phys. Rev. \textbf{D23}, 351 (1981).

\bibitem {diosi84}L. Diosi, G. Forgacs, B. Lukacs, and H. L. Frisch,
\emph{Metricization of thermodynamic-state space and the renormalization
group}, Phys. Rev. \textbf{A29}, 3343 (1984).

\bibitem {diosi96}L. Diosi, K. Kulacsy, B. Lukacs, and A. Racz,
\emph{Thermodynamic length, time, speed, and optimum path to minimize entropy
production}, J. Chem. Phys. \textbf{105}, 11220 (1996).

\bibitem {crooks17}G. M. Rostskoff, G. E. Crooks, and E. Vanden-Eijnden,
\emph{Geometric approach to optimal nonequilibrium control: Minimizing
dissipation in nanomagnetic spin systems}, Phys. Rev. \textbf{E95}, 012148 (2017).

\bibitem {andresen94}B. Andresen and J. M. Gordon, \emph{Constant
thermodynamic speed for minimizing entropy production in thermodynamic
processes and simulated annealing}, Phys. Rev. \textbf{E50}, 4346 (1994).

\bibitem {spirkl95}W.\ Spirkl and H. Ries, \emph{Optimal finite-time
endoreversible processes}, Phys. Rev. \textbf{E52}, 3485 (1995).

\bibitem {beretta05}E. P. Gyftopoulos and G. P. Beretta, \emph{Thermodynamics:
Foundations and Applications}, Dover Publications, Inc. (2005).

\bibitem {anandan90}J. Anandan and Y. Aharonov, \emph{Geometry of quantum
evolution}, Phys. Rev. Lett.\textbf{\ 65}, 1697 (1990).

\bibitem {cafaroPRA20}C. Cafaro, S. Ray, and P. M. Alsing, \emph{Geometric
aspects of analog quantum search evolutions}, Phys. Rev. \textbf{A102, }052607 (2020).

\bibitem {tolman48}R. C. Tolman and P. C. Fine, \emph{On the irreversible
production of entropy}, Rev. Mod. Phys. \textbf{20}, 51 (1948).

\bibitem {alsing19b}C. Cafaro and P. M. Alsing, \emph{Theoretical analysis of
a nearly optimal analog quantum search}, Physica Scripta \textbf{94}, 085103 (2019).

\bibitem {cafaroQR}C. Cafaro, S. Gassner, and P. M. Alsing, \emph{Information
geometric perspective on off-resonance effects in driven two-level quantum
systems}, Quantum Reports \textbf{2}, 166 (2020).

\bibitem {caves94}S. L. Braunstein and C. M. Caves, \emph{Statistical distance
and geometry of quantum states}, Phys. Rev. Lett. \textbf{72}, 3439 (1994).

\bibitem {sakurai94}J. J. Sakurai, \emph{Modern Quantum Mechanics},
Addison-Wesley Publishing Company, Inc. (1994).

\bibitem {brody03}D. C. Brody,\emph{ Elementary derivation for passage times},
J. Phys. \textbf{A}: Math. Gen. \textbf{36}, 5587 (2003)

\bibitem {gu21}P. M. Riechers and M. Gu,\emph{ Initial-state dependence of
thermodynamic dissipation for any quantum process}, Phys. Rev. \textbf{E103},
042145 (2021).

\bibitem {hayden21}P. Hayden and J. Sorce, \emph{On the magnitude of
dissipation in open quantum systems}, arXiv:quant-ph/2108.08316 (2021).

\bibitem {wu21}W. Wu and Z.-Z. Zhang, \emph{Controllable dynamics of a
dissipative two-level system}, Scientific Reports \textbf{11}, 7188 (2021).

\bibitem {brody19}D. C. Brody and B. Longstaff, \emph{Evolution speed of open
quantum dynamics}, Phys. Rev. Research\textbf{ 1}, 033127 (2019).

\bibitem {guarnieri21}E. O'Connor, G. Guarnieri, and S. Campbell, \emph{Action
quantum speed limits}, Phys. Rev. \textbf{A103}, 022210 (2021).

\bibitem {kittel58}C. Kittel, \emph{Elementary Statistical Physics}, John
Wiley \& Sons, Inc. (1958).

\bibitem {ramsey56}N. Ramsey, \emph{Thermodynamics and statistical mechanics
at negative absolute temperature}, Phys. Rev. \textbf{103}, 20 (1956).

\bibitem {purcell51}E. M. Purcell and R. V. Pound, \emph{A nuclear spin system
at negative temperature}, Phys. Rev. \textbf{81}, 279 (1951).

\bibitem {proctor57}A. Abragam and W. G. Proctor, \emph{Experiments on spin
temperature}, Phys. Rev. \textbf{106}, 160 (1957).

\bibitem {crooks12}G. E. Crooks, \emph{Fisher information and statistical
mechanics}, Technical note 008v4, http://threeplusone.com/sher (2012).

\bibitem {adolfo18}S. B. Nichols, A. del Campo, and J. R. Green,
\emph{Nonequilibrium uncertainty principle from information geometry}, Phys.
Rev. \textbf{E98}, 032106 (2018).
\end{thebibliography}
\end{document}